%% file: main.tex
\documentclass[fleqn,usenatbib]{mnras}

\usepackage{newtxtext,newtxmath}

\usepackage[T1]{fontenc}

\DeclareRobustCommand{\VAN}[3]{#2}
\let\VANthebibliography\thebibliography
\def\thebibliography{\DeclareRobustCommand{\VAN}[3]{##3}\VANthebibliography}

\usepackage{graphicx}	
\usepackage{amsmath}	
\usepackage{orcidlink}

\title[Investigating AME near G107.2+5.20 in Ku-band]{Investigating anomalous microwave emission near G107.2+5.20 in Ku-band with the Green Bank Telescope}

\author[J. E. Shroyer et al.]{
Jordan E. Shroyer\orcidlink{0000-0003-0514-9034},$^{1,2}$\thanks{E-mail: js6mu@virginia.edu}
Bradley R. Johnson\orcidlink{0000-0002-6898-8938},$^{1}$
Dillon J. Bass\orcidlink{0009-0001-1448-315X},$^{1}$
Anna Dignan\orcidlink{0009-0000-8877-135X},$^{1}$
\newauthor
Stuart E. Harper\orcidlink{0000-0001-7911-5553},$^{3}$
Brian S. Mason\orcidlink{0000-0002-8472-836X},$^{2}$
Andrey Moore\orcidlink{0009-0008-2816-9939}$^{1}$
and
Eric J. Murphy\orcidlink{0000-0001-7089-7325}$^{2}$
\\
$^{1}$Department of Astronomy, The University of Virginia, 530 McCormick Rd, Charlottesville, VA 22904 USA\\
$^{2}$National Radio Astronomy Observatory, 520 Edgemont Rd, Charlottesville, VA 22903 USA \\
$^{3}$ Jodrell Bank Centre for Astrophysics, Department of Physics \& Astronomy, The University of Manchester, Oxford Road, Manchester, M13 9PL, UK \\
}

\date{Accepted XXX. Received YYY; in original form ZZZ}

\begin{document}
\label{firstpage}
\pagerange{\pageref{firstpage}--\pageref{lastpage}}
\maketitle

\begin{abstract}
Anomalous microwave emission (AME) is 30 GHz-peaking continuum emission thought to arise from spinning dust grains. 
Observations suggest that the local environment shapes the AME spectral energy distribution (SED), so building a spatially resolved sample of AME regions is a key step towards understanding its emission mechanism. 
Using the Green Bank Telescope Ku-band receiver, we obtained a $\sim$1 arcmin resolution map of radius 1.25$^\circ$ centered on G107.2+5.20. 
Our first objective was to constrain the low-frequency side of the AME SED with 13 GHz data.
Using matched-resolution aperture photometry, we measure the SED from 408 MHz–3 THz and fit two emission models: one including spinning dust, the other optically thick free-free emission. 
We find that the spinning dust model is superior, with an amplitude of $14.1\pm1.1$ Jy and a peak frequency of $27\pm2$ GHz. 
Our second objective was to spatially locate excess 13 GHz emission consistent with spinning dust. 
We compare our Ku-band map to multi-wavelength gas and dust tracers at $\sim$4 arcmin resolution. 
We observe two sources of 13 GHz excess at $3\sigma$ significance consistent with spinning dust emission, though potential contributions from optically thick free-free emission remain ambiguous. ``Region A'' (G106.95+5.19) is spatially coincident with peak dust radiance. ``Region B'' (G107.08+4.90) is spatially coincident with peak PAH abundance. If regions A and B are indeed sources of spinning dust emission, they are a resolved example of how AME-dust correlation varies with environment.
\end{abstract}

\begin{keywords}
radiation mechanisms: general -- 
radio continuum: ISM -- 
ISM: dust, extinction -- 
ISM: H\,\textsc{ii} regions --
cosmic background radiation --
diffuse radiation

\end{keywords}



\section{Introduction}\label{sec:introduction}
Anomalous microwave emission (AME) is a component of diffuse Galactic emission observed between 10--60~GHz that cannot be attributed to synchrotron, free-free, or thermal dust emission. 
AME was first detected by cosmic microwave background (CMB) studies and observed to be spatially correlated with thermal dust emission \citep{kogut_microwave_1996, leitch_anomalous_1997}. 
AME is now widely recognized as an important component of Galactic emission--in the Galactic plane, it can account for up to 45~per cent of emission near its 30~GHz peak, constituting a significant low-frequency CMB foreground in temperature \citep{planck_collaboration_planck_2015}.
While observed AME polarization limits on the order of 1~per cent have been sufficient for past CMB studies, as new surveys probe ever-fainter sensitivity regimes even low levels of AME polarization could bias foreground estimation and compromise B-mode detection experiments \citep{gonzalez-gonzalez_quijote_2025, armitage-caplan_impact_2012, remazeilles_sensitivity_2016}.

The degree to which AME could bias future CMB studies is closely related to its emission mechanism.
AME is generally thought to arise from electric dipole radiation by nanometer-sized spinning dust grains with a non-zero electric dipole moment.
While other mechanisms have been proposed, they fail to individually account for all-sky trends, so, if present, they are likely sub-dominant to spinning dust (see \cite{dickinson_state--play_2018}, and references therein). 
However, it remains unclear exactly which small-grain population(s) give rise to spinning dust emission, and grain composition and dynamics directly affect whether polarized emission is likely or even possible \citep{hensley_simons_2022}. 
The open question of the AME ``carrier particle'' is also consequential for star formation studies because AME can constitute a significant fraction of 33 GHz continuum emission--a commonly used star formation rate (SFR) tracer--and a universally reliable AME tracer and SFR correction factor have not been identified.
Polycyclic aromatic hydrocarbons (PAHs) are a favored ``carrier particle'' due to their small size and high abundance, yet on degree scales, all-sky studies attempting to link AME to PAHs have produced mixed results.
Degree-scale observations by \cite{hensley_case_2016} find that 12 $\mu$m PAH emission traces AME no better than other dust tracers, while \cite{sponseller_statistical_2025} compare 3.3 $\mu$m PAH emission against thermal dust and find that although far-IR thermal dust emission is overall the best tracer, a substantial fraction (17--37 per cent) of the regions they consider are better traced by PAHs.
AME also seems to vary with environment \citep{cepeda-arroita_detection_2021, hensley_polycyclic_2022}. 
Studies using data at 30--60 arcmin resolution are limited by the fact that they blend many ISM phases within one photometry aperture. 
Therefore, to better understand the grain population(s) and environments that drive spinning dust emission, it is necessary to obtain high-resolution AME detections across a diverse set of regions. 
To date, this has been achieved for a limited sample of regions, including observations of $\lambda$ Orionis by \cite{harper_comap_2025}, $\rho$ Ophiuchus by \cite{casassus_resolved_2021}, LDN 1622 by \cite{harper_observations_2015} and \cite{casassus_morphological_2006}, various dark clouds by \cite{scaife_high-resolution_2010} and cold cores by \cite{tibbs_carma_2015}. 

In this study, we add to the sample of resolved observations of AME regions by performing arcminute resolution follow-up observations of a 1.25$^\circ$-radius region centered on G107.2+5.20.
We chose G107.2+5.20 because it is one of the few \textit{Planck}-identified AME targets that can be observed from the Green Bank Telescope in the Northern hemisphere \citep{planck_collaboration_planck_2014}.
At a distance of 910 pc, 1 arcmin resolution corresponds to roughly 0.25~pc, allowing us to resolve structures like molecular clouds and observe the region morphology at frequencies where AME can be directly detected. 
Like many other AME sources, the G107.2+5.20 region contains a variety of radiative environments and particles which are potential carriers of AME. 
Sharpless 140 (S140) is an H\,\textsc{ii} region which forms an interface with molecular cloud Lynds 1204 (L1204) \citep{sharpless_catalogue_1959, lynds_catalogue_1962}. 
There is also a large region of diffuse H\,\textsc{ii} \citep{abitbol_constraining_2018}.
Infrared observations have revealed several young stellar objects embedded in S140, as well as a protostellar disk and a large molecular outflow with an associated reflection nebula \citep{harvey_first_2012, maud_high_2013}.
Interferometric observations by 
\cite{perrott_investigating_2013} with the Arcminute Microkelvin Imager Small Array (AMI SA) and single-dish observations by \cite{abitbol_constraining_2018} with the Robert C. Byrd Green Bank Telescope (GBT) C-band receiver measured a radio SED consistent with either spinning dust or optically thick free-free emission, and the Q-U-I JOint Tenerife Experiment (QUIJOTE)  survey has reported a semi-significant AME detection with potentially substantial contributions from ultra-compact (UC) H\,\textsc{ii} emission \citep{poidevin_quijote_2023}.

In this work, we present new GBT Ku-band (13 GHz) observations that ``fill in'' the radio SED where the spinning dust signal is rising to better discriminate between the optically thick free-free and spinning dust scenarios. 
Furthermore, the GBT's arcminute resolution enables morphological comparison between different frequency maps of the region. 
If AME is localized, our observations are at a sufficient frequency and angular resolution to resolve it and observe spatial correlations with dust tracers on sub-degree scales.
Our study has two core objectives: (1) measure the rising edge of the AME SED to model the global emission mechanism, and (2) spatially locate any regions of excess 13 GHz emission that can be attributed to spinning dust or UC-H\,\textsc{ii}. To solidify our interpretation of the emission mechanism, we use multi-wavelength data at $\sim4$~arcmin resolution to trace gas and dust in the region. Our secondary objective is therefore to perform a limited, first-order analysis of the environment(s) in the region to distinguish between likely UC-H\,\textsc{ii} or spinning dust sources, taking into account the local dust grain populations.

The paper is organized as follows. Section \ref{sec:data} describes the data used in this study, outlining our new GBT Ku-band observations, our re-processing of the GBT C-band observations from \cite{abitbol_constraining_2018}, and the ancillary data sets used. 
Section \ref{sec:spectral-analysis} describes the aperture photometry, SED construction, and model fitting results. 
Section \ref{sec:morphological-analysis} describes the morphological analysis we performed to locate the source of the AME within our map and our comparisons to multi-wavelength gas and dust tracers. 
Section \ref{sec:discussion} is the discussion and Section \ref{sec:conclusions} is the conclusion.

\section{Data Acquisition and Preparation}\label{sec:data}
For this study we acquired new data in the Ku-band (12--18~GHz) using the Green Bank Telescope (GBT) with the VErsatile GBT Astronomical Spectrometer (VEGAS) backend to map a region of radius 1.25$^\circ$ centered on G107.2+5.20.
We also use existing GBT data in the C band (4--8 GHz) from \cite{abitbol_constraining_2018} and ancillary survey data spanning 408~MHz to 25~THz for spectral energy density (SED) analysis and modeling (Sec. \ref{sec:spectral-analysis}).
We convolve each map to a common resolution of 40~arcmin and perform background-subtracted aperture photometry to estimate the total spectral flux density (SFD) of the region in each band, then plot and fit emission models to the SED. 
Our GBT Ku-band observations are at a reasonable frequency to detect and locate spinning dust emission, which must have a rising SED between 12--18~GHz.
We leverage the arcminute resolution of the GBT to compare the multiwavelength morphology (Sec. \ref{sec:morphological-analysis}) of the region.
Additional maps spanning 100--3.3~$\mu$m were used for morphological comparison only.
For the morphological comparison, we convolve the maps to a common 4 arcmin resolution and compare features against the Ku-band map, focusing on regions of excess 13 GHz emission that are spatially correlated with dust tracers, as predicted by the spinning dust hypothesis. 
We also consider the possibility of contributions from optically-thick free-free emission.

The GBT/VEGAS was chosen because it is uniquely capable of providing maps of the size, angular resolution, and spectral resolution that our study required, with a filled aperture to sample large angular scales.
The morphological comparison requires resolution of a few arcminutes at several frequencies to precisely locate sources of excess 13 GHz emission and compare them to dust and gas features. 
The GBT resolution is about 1 arcmin at Ku-band (13~GHz).
On the other hand, our SED analysis requires that all of the maps share the same photometry aperture and background annulus, so our minimum map size is constrained by the lowest-resolution data set, from the \textit{COBE}-DIRBE survey (see Table \ref{tab:data-sets}). 
\cite{abitbol_constraining_2018} found that the SED results did not strongly depend on the size and location of the annulus; based on those conclusions we determined that a map of radius 1.25$^\circ$ would be sufficient for our study.

\subsection{GBT Data: Ku-band Observations}\label{ssec:data-collection}

\begin{table} 
    \caption{Definitions of the spectral windows (SPWs) used for each receiver.}
    \label{tab:spw-defs}
    \centering
    
    \textbf{(a) Ku-band}\\
    \input{tables/tab1a_spwDefinitions_kuband}

    \vspace{0.5em}
    \vspace{0.5em}
    
    \textbf{(b) C-band}\\
    \input{tables/tab1b_spwDefinitions_cband}

    \vspace{0.5em}
    {\raggedright\footnotesize
        \textbf{Note:} Subtable (b) is reproduced from \cite{abitbol_constraining_2018}. 
        The raw bandwidth is 1.5 GHz--the actual bandwidth used is 0.7 GHz. 
        The band center $\nu_c$ corresponds to the center of the selected band, $\nu_s$, not the raw band, $\nu_r$, and was chosen to avoid filter roll-off, SPW overlap, and RFI-heavy channels.
        We report the beam full-width at half-max for each SPW as $12.60'/\nu_c$ based on the GBT observer guide.
    \par}
\end{table}

We conducted 11.5 hours of on-the-fly (OTF) mapping observations and flux calibrations with the GBT on September 10 and 11, 2021.\footnote{Project code: AGBT21B\_241} 
We used $\sim9.5$ hours to make maps of radius 1.25 degrees centered on G107.2+5.20, and used $\sim2$ hours for flux calibrations on 3C295. 
The Ku-band receiver is equipped with a noise diode which we switch continuously on low power during the observations at a rate of 25 Hz.
We use the 3C295 observations to calibrate the noise diode temperature, which we assume is stable over the course of the mapping observations. 
We then use the noise diode to calibrate the time-ordered mapping data.
More details on calibration are given in Sec. \ref{ssec:calibration}.
We chose the OTF daisy scan strategy (see Fig. \ref{fig:kumap-with-noise}) because the cross-linking enables the destriping mapmaking algorithm we used to more optimally remove low-frequency ``1/f'' noise from atmospheric fluctuations.
The map data was taken in five-minute intervals.
We used the maximum allowed mapping speed, $\sim35$~arcmin per second.
More details on mapmaking are given in Sec. \ref{ssec:mapmaking}.
Although we are interested in measuring the continuum emission, we used VEGAS in spectral line mode at the maximum 92 kHz resolution. 
We did this to enhance our ability to mitigate radio-frequency interference (RFI). 
The high spectral resolution and fast sample rate VEGAS offers allowed us to excise RFI in specific channels and integrations instead of averaging it into our raw data. 
More details on RFI excision are given in Sec. \ref{ssec:data-selection-rfi-excision}.
The GBT Ku-band receiver has two beams; however, we only calibrated and used data from the primary beam. 
VEGAS has four spectral windows (SPWs) for one beam, labeled A-D. 
The center frequencies and widths of the SPWs are given in Table \ref{tab:spw-defs}.
Ultimately, we omitted used SPW A from our analysis due to RFI corruption of the calibration scans, and we omitted SPWs C and D due to poor signal-to-noise.

\subsection{GBT Data: Band Selection and RFI Excision}\label{ssec:data-selection-rfi-excision}

\begin{table}
    \caption{Summary of flags from the spectral kurtosis (SK) and noise-to-signal ratio (NSR) cuts.}
    \label{tab:rfi_summary}
    \centering
    
    \textbf{(a) Ku-band}\\
    \input{tables/tab2a_FlagSummaryResults_kuband}

    \vspace{0.5em}
    
    \textbf{(b) C-band}\\
    \input{tables/tab2b_FlagSummaryResults_cband}

    \vspace{0.5em}
    {\raggedright\footnotesize
        \textbf{Note:} ``Channels flagged'' is the average per cent of channels flagged across all scans. ``Scans $>$1\% flagged'' is the fraction of scans with $>$1 per cent of channels flagged. Of the flags applied, the SK and NSR columns indicate the percent applied by each test--their sum may exceed 100 per cent because some spectral channels were flagged by both tests. SPW A is omitted because the calibration data contained excessive RFI.
    \par}
\end{table}

Radio-frequency interference (RFI) can be an issue in the Ku-band. 
We chose our SPW center frequencies and bandwidths ($\nu_c$ and $\Delta\nu_s$ in Table \ref{tab:spw-defs}, respectively) to avoid RFI, avoid overlap, and trimmed the filter roll-off at the band edges.
We used VEGAS at the maximum spectral resolution (mode 2, 92 kHz) so that we could identify and excise individual spectral channels corrupted by RFI.
Within the selected bandwidth $\Delta\nu_s$, we identified RFI-corrupted spectral channels using two standard tests: coefficient of variation and spectral kurtosis \citep{nita_generalized_2010, nita_statistical_2019}.
We used the same criteria in \cite{abitbol_constraining_2018} and \cite{walters_axions_2024}, 
and summarize the procedure here.
The coefficient of variation, or inverse signal-to-noise ratio (NSR$_\nu$) is used to identify channels with persistently high noise levels characteristic of RFI. 
For time-ordered spectral data $\xi_{\nu,t}$, where the subscript $\nu$ denotes frequency and the subscript $t$ denotes time,
\begin{equation}
    \mathrm{NSR}_\nu = \frac{\sigma_\nu}{\mu_\nu} = \frac{\sqrt{\langle (\xi_{\nu,t} - \mu_\nu)^2\rangle}_t}{\langle \xi_{\nu,t}\rangle_t}.
\end{equation}
We masked channels with NSR$_\nu$ greater than~7.5 times the median absolute deviation of the NSR$_\nu$.
The spectral kurtosis, $K_\nu$, is used to identify channels with non-Gaussian noise properties; RFI tends to be periodic and/or intermittent while astrophysical signals tend to be Gaussian. 
\begin{equation}
    K_\nu = \frac{\langle(\zeta_{\nu,t}-\mu_\nu)^4\rangle_t}{\langle(\zeta_{\nu,t}-\mu_\nu)^2\rangle_t^2}
\end{equation}
We chose to mask spectral channels with kurtosis greater than~7.5 times the times the median absolute deviation of $K_\nu$.

We apply these cuts first to the calibrator data and mask any spectral channels above the NSR$_\nu$ or $K_\nu$ threshold. 
Because RFI is typically polarized, we flag the LL and RR spectrometer outputs separately, then combine them into a single mask to be used on Stokes I = (LL + RR)/2.
We combine the flag masks with a logical OR--any channel that is flagged by either test in LL or RR is flagged in the combined mask.
We then apply the same threshold cuts to each five-minute mapping scan independently.
For a given mapping scan, we mask any channel that was flagged in either the calibrator scan or in that mapping scan.
For SPW B, 99 per cent of scans had less than 1 per cent of channels flagged. 
Of the flagged channels in SPW B, 82 per cent were flagged based on the Kurtosis cut and 46 per cent were flagged based on the NSR cut (some channels were flagged by both cuts).
We did not attempt to calibrate the data from SPW A because the calibration scans were excessively corrupted by RFI.

\subsection{GBT Data: Calibration and Time-Ordered Data Reduction}\label{ssec:calibration}

\begin{figure}
    \centering
    \includegraphics[width=\columnwidth]{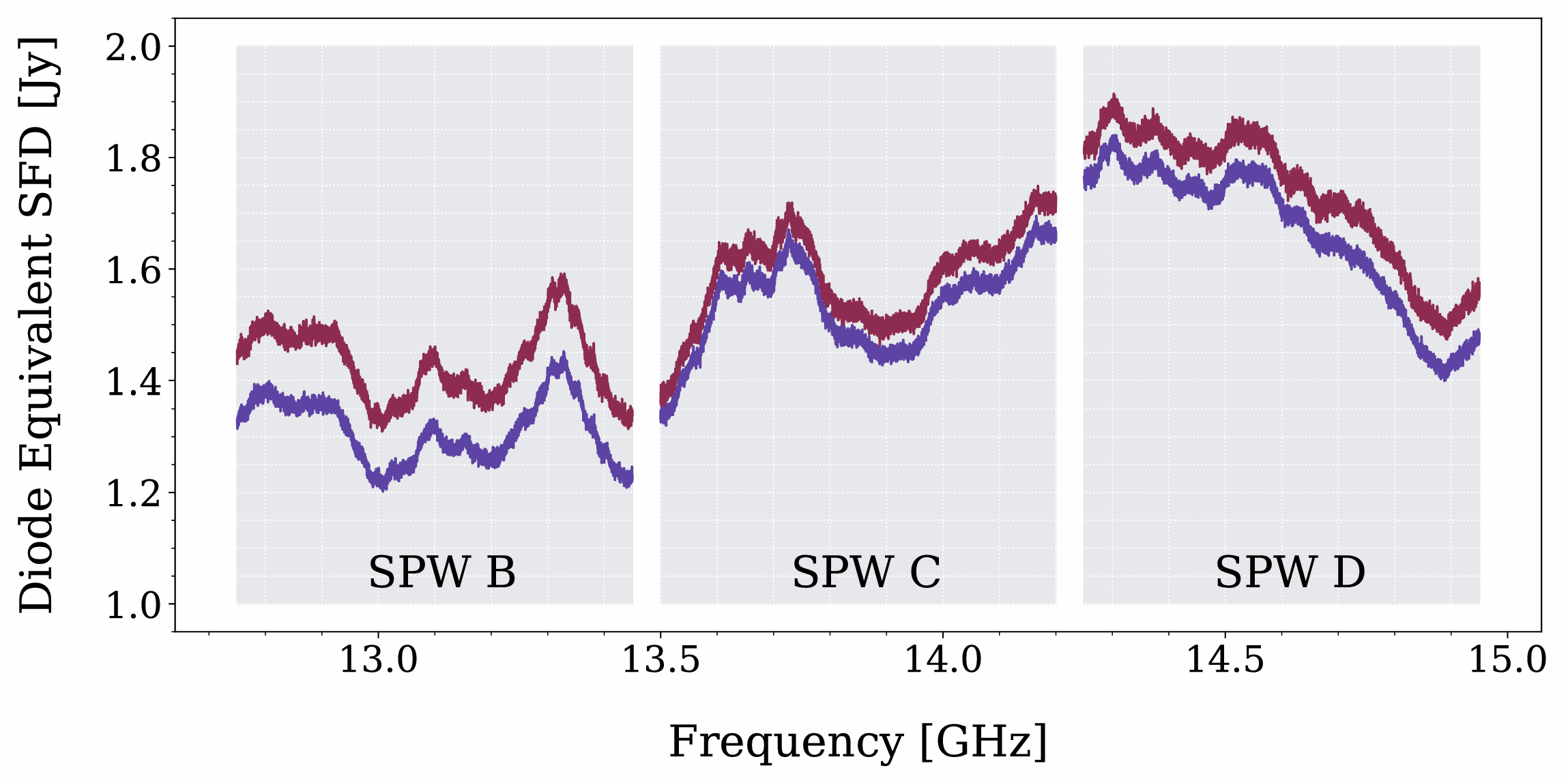}
    \caption{Noise diode spectrum, calibrated to units of Jy using astrophysical calibrator 3C295. The spectral window (SPW) bandwidths from Table \ref{tab:spw-defs}, indicated by the gray boxes, were selected to avoid known RFI sources; however, SPW A was omitted due to excessive RFI corruption in the calibration scans. Data were taken during two consecutive nights. The blue (bottom) spectrum is from the first night, and the red (top) spectrum is from the second.}
    \label{fig:noise-diode-spectrum}
\end{figure}

We wrote a custom data calibration and reduction pipeline in Python to prepare the data for mapmaking.
Observations consisted of two-minute on- and off-source flux calibration scans of 3C295, and a series of five-minute mapping scans centered on G107.2+5.20.
We define $x$ as the off-source calibration data, $y$ as the on-source calibration data, and $z$ as the mapping data.
Superscripts indicate whether the noise diode is on or off, and subscripts indicate a particular frequency channel and time.
So, $x_{\nu,t}^\text{on}$ is off-source calibration data at time $t$, spectral channel $\nu$, with the noise diode on. 

We need to estimate the system gain, $G_\nu$, to calibrate the time-ordered mapping data from arbitrary ADC ``counts'' to flux units (Jy). 
The receiver has active components that can vary on timescales shorter than our several-hour observing runs.
We use our flux calibration observations to measure the spectrum of the receiver's noise diode, which injects a signal into the readout chain between the antenna and the first active components and is stable on timescales longer than our observing run.
Then, we use the calibrated noise diode spectrum to calibrate each of our 5-minute mapping scans.
The noise diode was flashed on and off at 25 Hz during the calibration and mapping observations, so that every other measurement is effectively a measurement of its temperature. 

We calibrate the noise diode spectrum using the well-known astrophysical calibrator 3C295. 
We measure the time-averaged noise diode level $D_\nu$ in a spectral channel $\nu$ using our off-source calibration data with the diode switching on and off:
\begin{equation}
    D_\nu = \langle x_{\nu,t}^\text{on} - x_{\nu,t}^\text{off}\rangle_t.
\end{equation}
The time-averaged 3C295 level $S_\nu$ in spectral channel $\nu$ is calculated from pairs of on-off pointings with the diode off:
\begin{equation}
    S_\nu = \langle y_{\nu,t}^\text{off} - x_{\nu,t}^\text{off} \rangle_t.
\end{equation}
We average $D_\nu$ and $S_\nu$ over the duration of the on-off calibration scans, then obtain the calibrated noise diode signal, $P_\nu$, in units of Janskys using
\begin{equation}
    P_\nu = \frac{I_\nu}{S_\nu} D_\nu,
\end{equation}
where $I_\nu$ is the known SFD of 3C295 \citep{perley_accurate_2017}.
Using the diode signal, $P_\nu$, which we assume to be stable over the course of our observing run, we can now calibrate the RFI-cleaned mapping data, $z$.
For each scan, we calculate the average inverse receiver gain:
\begin{equation}
    G_\nu = \frac{P_\nu}{\langle z_{\nu,t}^{on} - z_{\nu,t}^{off}\rangle_t}.
\end{equation}
For a given scan, each integration in the time-ordered mapping data is calibrated using the $G_\nu$ for that scan and band-averaged according to the equation
\begin{equation}
    d_t = \langle \frac{G_\nu z_{\nu,t}^{off}}{\Omega_\nu} \rangle_\nu.
\end{equation}
To obtain units of Jy~Sr$^{-1}$, we divide by the beam solid angle, $\Omega_\nu$, defined as
\begin{equation}
    \Omega_\nu = \frac{\pi}{4 \log{2}} \text{FWHM}_\nu^2 
    \label{eq:fwhm_to_solidAngle}
\end{equation}
for diffuse emission, using the FWHM given in Table \ref{tab:spw-defs}.
For each SPW, we average over the selected bandwidth, $\Delta\nu_s$, given in Table \ref{tab:spw-defs}. 
These calibrated, band-averaged, time-ordered data are the inputs to our mapmaking algorithm, along with the telescope pointing coordinates for each $z_t$.

\subsection{GBT Data: Mapmaking}\label{ssec:mapmaking}

\begin{figure}
    \centering
    \includegraphics[width=0.9\columnwidth]{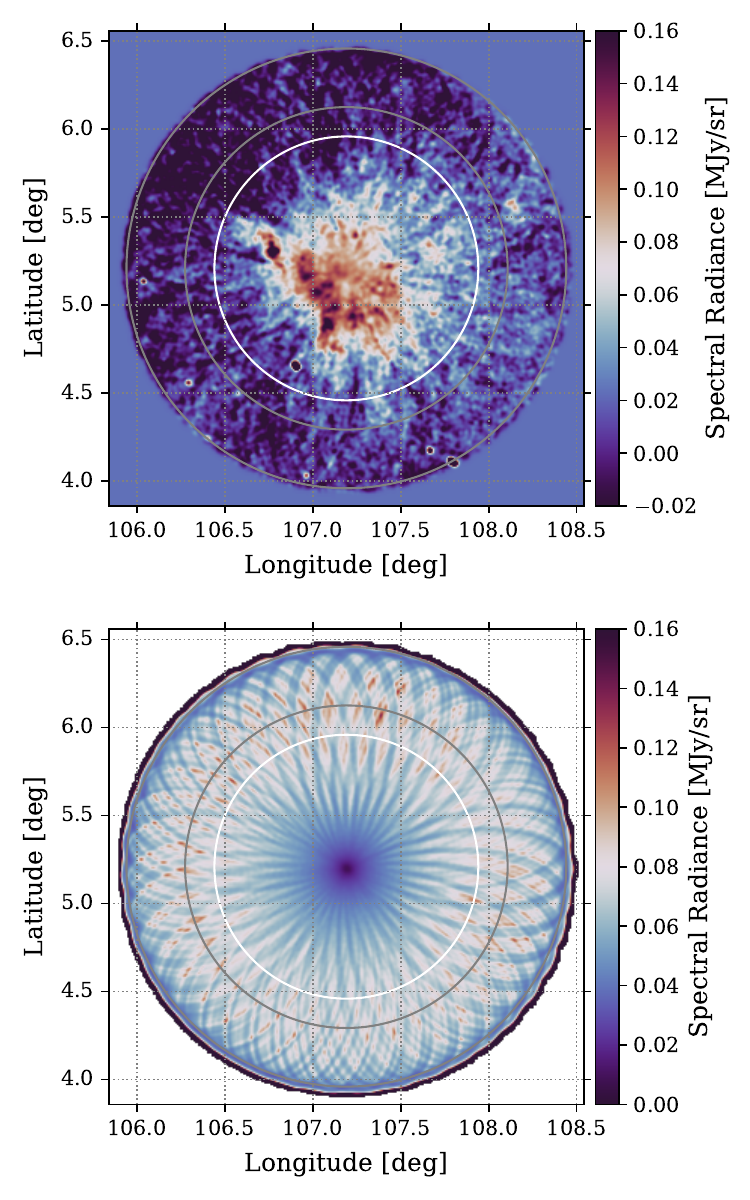}
    \caption{Top row: Intensity map from this study centered on G107.2+5.20 taken with the GBT Ku-band receiver, SPW B (13.1 GHz). 
    The aperture (white) and background annulus (gray) used for aperture photometry (sec. \ref{ssec:aperture-photometry}) are overlaid for reference. 
    Maps were constructed using the destriping algorithm described in Sec. \ref{ssec:mapmaking}, smoothed with a Gaussian beam using the FWHM for each SPW (Table \ref{tab:spw-defs}), and background-subtracted using the median value in the annulus. 
    Bottom: Noise map computed by the destriping algorithm.
    The ``petals'' are from the daisy scan pattern--one five-minute scan traces out $\sim5.5$ ``petals''.
    The center is more densely sampled, so the noise is lower at the center and higher in the edges of the map. 
    }
    \label{fig:kumap-with-noise}
\end{figure}

We made maps from the band-averaged time-ordered data (TOD), $d$, using the destriping algorithm described in \cite{harper_single-dish_2016}. 
The algorithm is based on the method described in \cite{delabrouille_analysis_1998},  \cite{sutton_map_2009}, and \cite{sutton_fast_2010}, and our usage is presented in \cite{abitbol_constraining_2018} and summarized here. 
A simple bin-and-average mapmaking approach works when the TOD have Gaussian (uncorrelated) noise; destriping becomes necessary when atmospheric fluctuations and variations in the receiver gain and system temperature give rise to correlated ``1/f'' noise.
In the TOD, 1/f noise creates low-frequency baseline drifts.
In the maps, 1/f noise creates stripes along the telescope scanning path if a naive bin-and-average approach is used.
The destriper works by modeling the calibrated TOD, $d$, as combination of the true sky signal ($Pm$), white noise ($n_w$), and 1/f noise ($Fa$).
We model the calibrated TOD (length $n_\text{samples}$) from one SPW as 
\begin{equation}
    d = Pm + n_w + Fa.
    \label{eq:data-model}
\end{equation}
The aim is to recover the map vector, $m$ of the true sky signal (length $n_\text{pix}$). $P$ is the pointing matrix ($n_\text{samples}\times n_\text{pix}$) which transforms on-sky pixel locations to time positions in the TOD. $n_w$ is the white noise vector. $Fa$ is the assumed model for the correlated 1/f noise. 
The 1/f noise is modeled as a series of linear offsets of varying amplitude ($a$) where the length (related to $F$) in the TOD is a user-selected parameter and the amplitudes are solved for using a least-squares minimization routine. 
The 1/f noise model is then subtracted from the TOD to obtain the maximum-likelihood map estimate
\begin{equation}
    m = (P^T N^{-1}P)^{-1} P^T N^{-1}(d-Fa),
\end{equation}
 This is essentially a noise-weighted histogram--it is assumed to be sky signal plus white noise and is used to make the map shown in Fig. \ref{fig:kumap-with-noise}.
The mapmaker also computes an uncertainty-per-pixel, which is shown in the bottom panel of Fig. \ref{fig:kumap-with-noise}.
Finally, the maps are smoothed using Gaussian convolution.
For aperture photometry we do Gaussian convolution to 40' (eq. \ref{eq:common-beam-smoothing}), and for the morphological analysis we convolve to 4' to match the C-band (4~GHz) data.

\begin{table*}
    \centering
    \caption{Summary of the data sets used in this study.}
    \label{tab:data-sets}
    \input{tables/tab3_dataSetSummary}
    \vspace{0.5em}
    \centering
    
    \footnotesize *used for morphological comparison only; $^+$omitted from the SED fitting, aperture correction applied;
    $^\dagger$omitted from SED fitting due to CO line emission.
\end{table*}

\subsection{Ancillary Data: Radio, Sub-Millimeter, and Infrared}\label{ssec:ancillary-data}

The SED analysis requires spectral coverage from the radio to the far-IR.
The morphological comparison requires spatial resolution of a few arcminutes at wavelengths that trace gas and dust in the ISM.
The additional data sets we used were, in order of low to high frequency: the Canadian Galactic Plane Survey (CGPS) \citep{taylor_canadian_2003,tung_high_2017, landecker_survey_2010}, the Stockert northern sky survey \citep{reich_radio_1982,reich_radio_1986}, GBT C-band maps \citep{abitbol_constraining_2018},
selected bands fromthe Q-U-I JOint Tenerife Experiment (QUIJOTE) \citep{rubino-martin_quijote_2023},
\textit{Planck} LFI and HFI maps \citep{planck_results_2016a}, and \textit{COBE}-DIRBE \citep{hauser_cobe_1998}. 
For our morphological comparison, we also used the improved reprocessing of the IRAS survey (IRIS) \citep{miville-deschenes_iris_2005} and SPHEREx data \citep{spherex_team_spherex_2025}.
Table \ref{tab:data-sets} lists the frequency, resolution, our measured aperture SFD (when applicable), and key references for each data set used in this study. 
This section describes each ancillary data set used and any color corrections and unit conversions we applied.

\subsubsection{Ground-based Radio Data}
The Canadian Galactic Plane Survey (CGPS)\footnote{The CGPS data are available online at \url{https://www.cadc-ccda.hia-iha.nrc-cnrc.gc.ca/en/cgps/}} \citep{taylor_canadian_2003, tung_high_2017} provided the lowest-frequency data used in our study-- 408 and 1420~MHz. 
To capture small- and large-scale structure, the CGPS combined data from the Dominion Radio Astrophysical Observatory (DRAO) Synthesis Telescope with single-dish data. 
The 408 MHz single dish data was provided by the $51$~arcmin resolution Haslam all-sky continuum survey \citep{haslam_408_1981, haslam_408-mhz_1982}.
The 1420 MHz single dish data was provided by the $35$~arcmin resolution Stockert northern sky survey \citep{reich_radio_1982, reich_radio_1986}.
The DRAO synthesis telescope gives the CGPS maps a resolution of 3 and 1 arcmin at 408 and 1420 MHz, respectively, which makes them useful for morphological comparison and identifying radio point sources.
The 408 MHz CGPS map fully covers our region of interest, so it was used for both morphological comparison and aperture photometry.
The 1420~MHz CGPS map was used for morphological comparison only; with a maximum galactic latitude of only $+5.6^{\circ}$, it excludes roughly one third of the $1.25^{\circ}$-radius region around G107.2+5.20 we used for aperture photometry.
Instead, the 1420~MHz aperture photometry measurement was made using data from the aforementioned Stockert 21~cm Northern sky survey\footnote{The all-sky 21~cm data are available online at \url{https://www3.mpifr-bonn.mpg.de/survey.html}} alone \citep{reich_radio_1982, reich_radio_1986}.
The CGPS and Stockert data are reported in units of brightness temperature, so we converted them to flux units (Jy sr$^{-1}$) using the Rayleigh-Jeans approximation:
\begin{equation}
    I = \frac{2 \nu^2 k_\text{B}}{c^2} T_\text{B} \times 10^{26}.
\label{eq:rayleigh-jeans}
\end{equation}

From 4-8~GHz, we used the GBT C-band (4-8 GHz) data that members of our team analyzed in \cite{abitbol_constraining_2018}. 
This work uses the same calibration, data reduction, and mapmaking procedure.
However, an updated data reduction and calibration pipeline was written in Python 3 for this study, so it was used to re-process the C-band data alongside the new Ku-band data. 
The aperture photometry results from the re-processed C-band maps are consistent with the ones presented in \cite{abitbol_constraining_2018}.

We used the 17 and 19~GHz bands from the QUIJOTE mid-frequency instrument (MFI) wide survey in our SED analysis \citep{rubino-martin_quijote_2023}. 
We omit the 11 and 13 GHz QUIJOTE data from the modeling due to their $\sim55$ arcmin resolution, coarser than the minimum 41 arcmin resolution on which our GBT map size was based. 
However, we do plot the 11 and 13 GHz points on our SED after applying an estimated aperture correction for comparison with our GBT results, finding that the QUIJOTE and GBT measurements are consistent within photometric errors. 

\subsubsection{Millimeter and Sub-Millimeter Satellite Data}

We used \textit{Planck} data for 30-857 GHz\footnote{The \textit{Planck} data is available on the \textit{Planck} Legacy Archive}.
We use the 2015 public release 2 (PR2) data because it is better optimized for foreground temperature studies compared to the 2018 PR3 data release, which is better optimized for polarization studies \citep{akrami_planck_2020}.
The \textit{Planck} 353 GHz data in particular was useful for our morphological comparison because it has a relatively high ($4.8'$) resolution and is a good tracer for thermal dust emission.
Below 545~GHz, the \textit{Planck} PR2 data are given in units of $K_{CMB}$ and must be converted to flux units.
We perform the unit conversion from K$_\text{CMB}$ to MJy~sr$^{-1}$ and color-corrections for the high-frequency data according to the \textit{Planck} Legacy Archive (PLA) explanatory supplement using the tables provided\footnote{modified blackbody dust model, $T_\text{D}$=21~K, $\beta_\text{D}$=1.48}.
We omit the 100 and 217 GHz observations from our SED fitting because they are biased by molecular CO lines. 

DIRBE was the highest-frequency data set used for our SED fitting, up to 3 THz. 
We did not use higher frequencies because the dust absorption and emission becomes more complex beyond 3 THz.
DIRBE was the lowest-resolution survey included in our data set at the time the GBT observations were obtained, so it set our aperture size and the common beam resolution for the aperture photometry. 
The DIRBE documentation reports the beam solid angle; for consistency with the other experiments, we use Equation \ref{eq:fwhm_to_solidAngle} to convert to the corresponding beam FWHM.
We applied color corrections using the DIRBE explanatory supplement table and assumed a dust temperature of 20~K and a spectral index of $\beta_D=1.5$. 

\subsubsection{Infrared satellite data}
Finally, for the morphological analysis, we used images from IRIS (100 \textendash 12 $\mu$m) and derived PAH emission maps from SPHEREx.
The IRIS images have comparable resolution to the GBT C-band data and can be used to trace dust emission and abundance from grains of various sizes.
The SPHEREx maps were made using the latest public data release (QR2) following the methods of \cite{cukierman_spectral_2026} and \cite{murgia_spherex_2026}, and are further described in Section \ref{ssec:planck-dust-params}.

\section{Aperture Photometry and SED Modeling}\label{sec:spectral-analysis}

\subsection{Aperture Photometry}\label{ssec:aperture-photometry}

\begin{figure*}
    \centering
    \includegraphics[width=\textwidth]{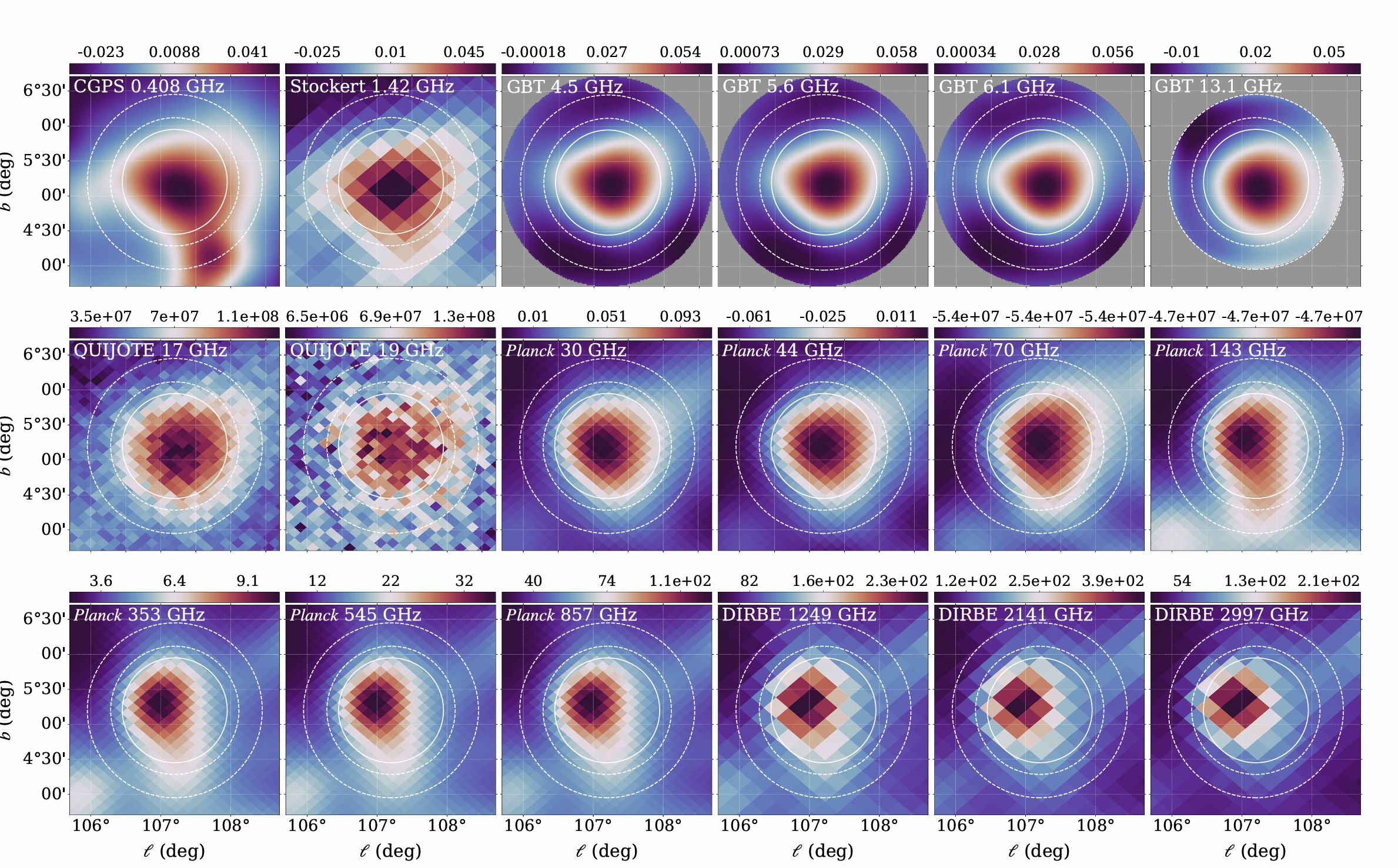}
    \caption{
        Maps used for matched-resolution aperture photometry and SED analysis described in Section \ref{sec:spectral-analysis}. 
        Each map was convolved with the 41' DIRBE beam, the coarsest resolution survey in our data set. 
        The solid white circle denotes the photometry aperture, and the dashed white circles denote the background annulus. 
        Each map was background-subtracted using the median value in the background annulus.
        All units are in MJy sr$^{-1}$. 
    }
    \label{fig:apphot-maps}
\end{figure*}

\begin{table*}
    \caption{Summary of foreground models, free parameters, and fixed model parameters.}
    \label{tab:foreground-models}
    \centering
    \input{tables/tab4_modelParamSummary}
\end{table*}

We use matched-resolution aperture photometry to measure the spectral flux density (SFD) in each band and construct a spectral energy distribution (SED) for the region. 
We use 18 total maps spanning 408 MHz to 3 THz.
See Table \ref{tab:data-sets} for details.
We first smooth each map to a common resolution of 41~arcsec by convolving with a Gaussian beam,
\begin{equation}
    \text{FWHM} = \sqrt{(41')^2 - \theta^2},
    \label{eq:common-beam-smoothing}
\end{equation}
where $\theta$ is the beam FWHM of the map at native resolution and 41~arcmin is the resolution of the largest DIRBE beam.
The common-resolution maps with photometry apertures and annuli are shown in Figure \ref{fig:apphot-maps}.
We chose to use a photometry aperture of radius $45$ arcmin and a background annulus with radii $55-75$ arcmin.
For each map, we measure the spectral flux density (SFD) by summing the pixels inside the aperture after converting to units of Jy pixel$^{-1}$.
We estimate the local background to provide a relative zero-point using the median pixel value in the annulus. 
The annulus dimensions are constrained by the GBT Ku-band maps, which have the smallest spatial extent of any in our data set. 
\cite{abitbol_constraining_2018} 
made C-band maps of radius 1.5$^\circ$ but found that their results did not depend strongly on annulus position, so we chose to make smaller maps of radius 1.25$^\circ$ maps in Ku-band to be efficient with observing time.
We also use the annulus to estimate the standard error per pixel by estimating the standard error of the median computed across the annulus pixels, adjusted to account for pixels with correlated noise: 
\begin{equation}
    \sigma = n_\text{pix}\times\sigma_\text{ann}\sqrt{\frac{1}{n_\text{ap}} + \frac{\pi}{2}\frac{1}{n_\text{ann}}}.
\end{equation}
See equations (4) and (5) in \cite{genova-santos_quijote_2015} for further details. 
The SFD and uncertainty measurements are reported in Table \ref{tab:data-sets}. The SED and foreground models are shown in Figure \ref{fig:seds-and-fits}.

\subsection{Foreground Modeling}\label{ssec:foreground-models}

Previous work has established that the G107.20+5.20 region exhibits excess 30~GHz emission that cannot be attributed to thermal dust, optically thin free-free, or synchrotron emission; however, whether this excess arises from spinning dust, optically thick free-free emission, or a combination remains ambiguous \citep{perrott_investigating_2013, abitbol_constraining_2018, poidevin_quijote_2023}.
We compare emission models that include either a spinning dust component or an optically-thick free-free component and report the median parameters and errors obtained from MCMC sampling.
First, we model the excess as AME using a four-component model including spinning dust, 
\begin{multline}
    S(\nu) = S_\text{SD}(\nu, A_\text{SD}, \nu_\text{SD}) 
    + S_\text{FF}(\nu, \text{EM}) \\ 
    + S_\text{D}(\nu, A_\text{D}, \beta_\text{D}, T_\text{D})
    + S_\text{CMB}(\nu, A_\text{CMB}).
    \label{eq:spinning-dust-model}
\end{multline}
Second, we model the excess as optically thick free-free emission using a four-component model including ultra compact (UC) H\,\textsc{ii}
\begin{multline}
    S(\nu) = S_\text{UCHII}(\nu, \text{EM}_\text{UCHII},\theta_\text{UCHII}) 
    + S_\text{FF}(\nu, \text{EM}) \\ 
    + S_\text{D}(\nu, A_\text{D} , \beta_\text{D}, T_\text{D})
    + S_\text{CMB}(\nu, A_\text{CMB}).
    \label{eq:uchii-model}
\end{multline}
We do not include synchrotron emission because the SED is flat at low frequencies.
We convert models from units of K to Jy sr$^{-1}$ using the Rayleigh-Jeans equation (eq. \ref{eq:rayleigh-jeans}).
We list our models, free parameters, and fixed parameters in Table \ref{tab:foreground-models}. 
Each model has a total of seven free parameters.

We model spinning dust using the phenomenological log-normal model proposed by \cite{stevenson_derivation_2014}.
The free parameters are the peak frequency $\nu_\mathrm{SD}$ and amplitude $A_\mathrm{SD}$ of the parabola at $\nu_\mathrm{SD}$. Our data cannot robustly constrain the dimensionless width, $W_\mathrm{SD}$ of the parabola, so we used a fixed width of 0.7--the upper end of the expected range of values for Galactic AME sources reported by \cite{cepeda-arroita_detection_2021}.

Free-free emission, or thermal bremsstrahlung radiation, is produced by free electrons scattering off of ions. 
It is the dominant emission component in the CGPS 408 and 1420 MHz and GBT 4-6 GHz maps presented in Section \ref{sec:morphological-analysis}.
Most of the free-free emission in our region comes from the H\,\textsc{ii} region S140 and from the large diffuse cloud near the center of the field.
We use the brightness temperature model presented in  \cite{2011piim.book.....D}.

For optically thick free-free emission, we modify the model for diffuse free-free emission.
Ultra-compact H\,\textsc{ii} regions have characteristically small spatial extents, with diameters on the order of 0.1 pc. 
Hyper-compact H\,\textsc{ii} regions can be even smaller and denser, with diameters $<0.01$~pc. 
At a distance of 910 pc and angular resolution of 1 arcmin, features smaller than 0.26~pc are unresolved. 
Therefore, we modify our free-free emission model, multiplying by a solid angle parameter to account for the unknown size of an unresolved source of compact H\,\textsc{ii}.

Thermal emission from heated dust grains is the dominant emission component above $\sim100$~GHz.
We model the thermal dust emission as a modified blackbody curve with a power-law emissivity. 
We use a reference frequency of 545 GHz and fit the dust amplitude $A_\text{D}$, dust temperature $T_\text{D}$, and dust spectral index $\beta_\text{D}$ as free parameters. 
Some degeneracy between $T_\text{D}$ and $\beta_\text{D}$ is expected.

Because of the CMB temperature anisotropy, the CMB temperature varies between our aperture and annulus.
We account for this by modeling the CMB contribution to our SED as the first derivative of a blackbody with respect to temperature. 
The amplitude of the derivative is a free parameter.
We considered using CMB-subtracted \textit{Planck} frequency maps, but chose not to because our source is near the Galactic plane where the local CMB solution depends on foreground models and priors. 

\subsection{SED fits and MCMC analysis}\label{ssec:mcmc}

\begin{figure}
    \centering
    \includegraphics[width=\linewidth]{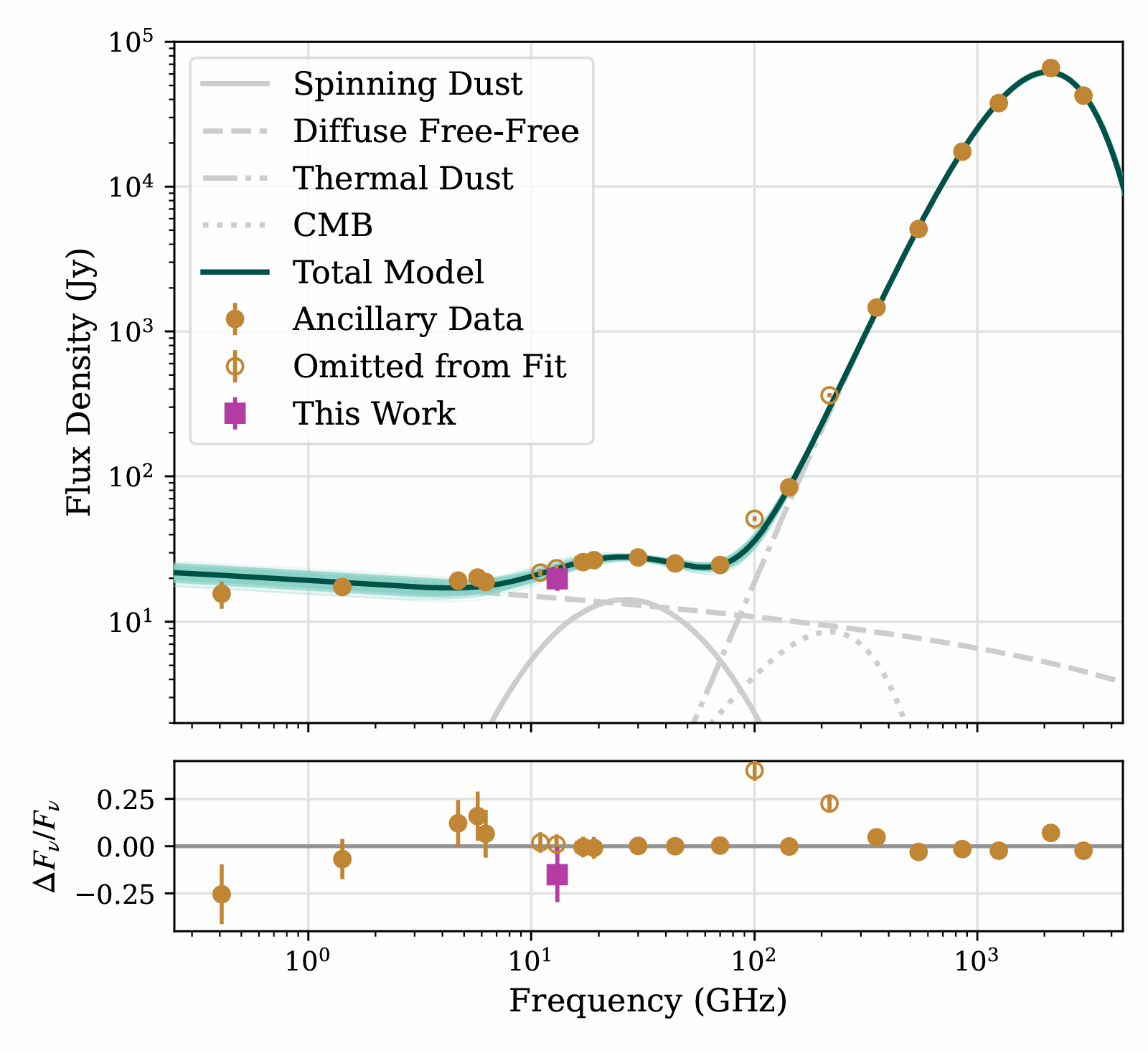}
    \includegraphics[width=\linewidth]{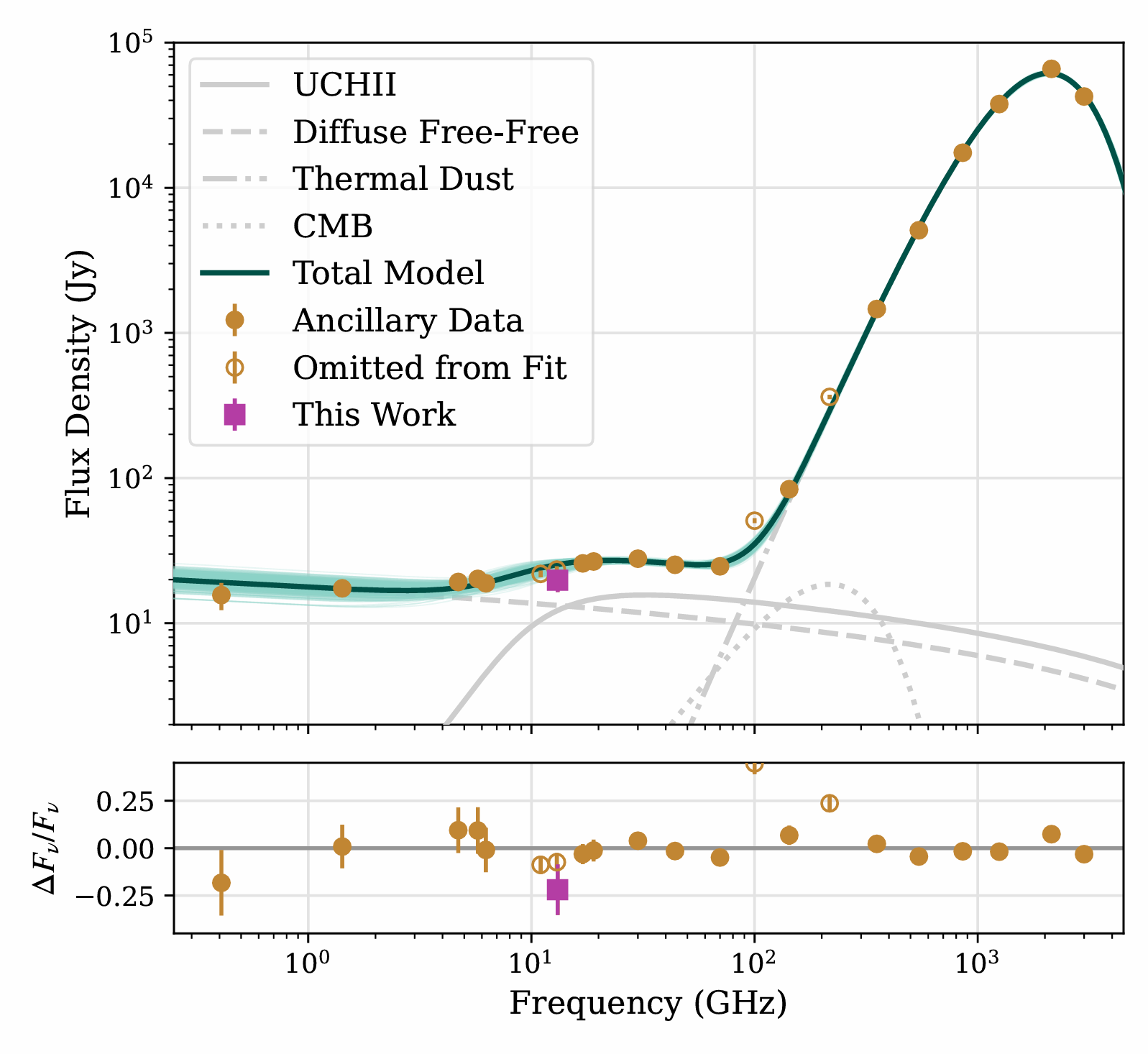}
    \caption{SEDs for G107.2+5.20 fit using a spinning dust component (top) versus a UCH\,\textsc{ii} component (bottom). In the bottom panel, we have plotted the absolute value of the CMB component. The pink square indicates the measurement from the GBT Ku-band maps produced for this study. 
    We plot the aperture-corrected 11 and 13 GHz QUIJOTE data with open circles, but exclude them from the fit due to coarse resolution.
    We plot the 100 and 217 GHz \textit{Planck} data with open circles, but exclude them from the fit due to contamination by CO line emission. The SED modeled with a spinning dust component is more consistent with our Ku-band data than the SED modeled with a UCH\,\textsc{ii} component.}
    \label{fig:seds-and-fits}
\end{figure}
\begin{figure*}
    \includegraphics[width=0.85\textwidth]{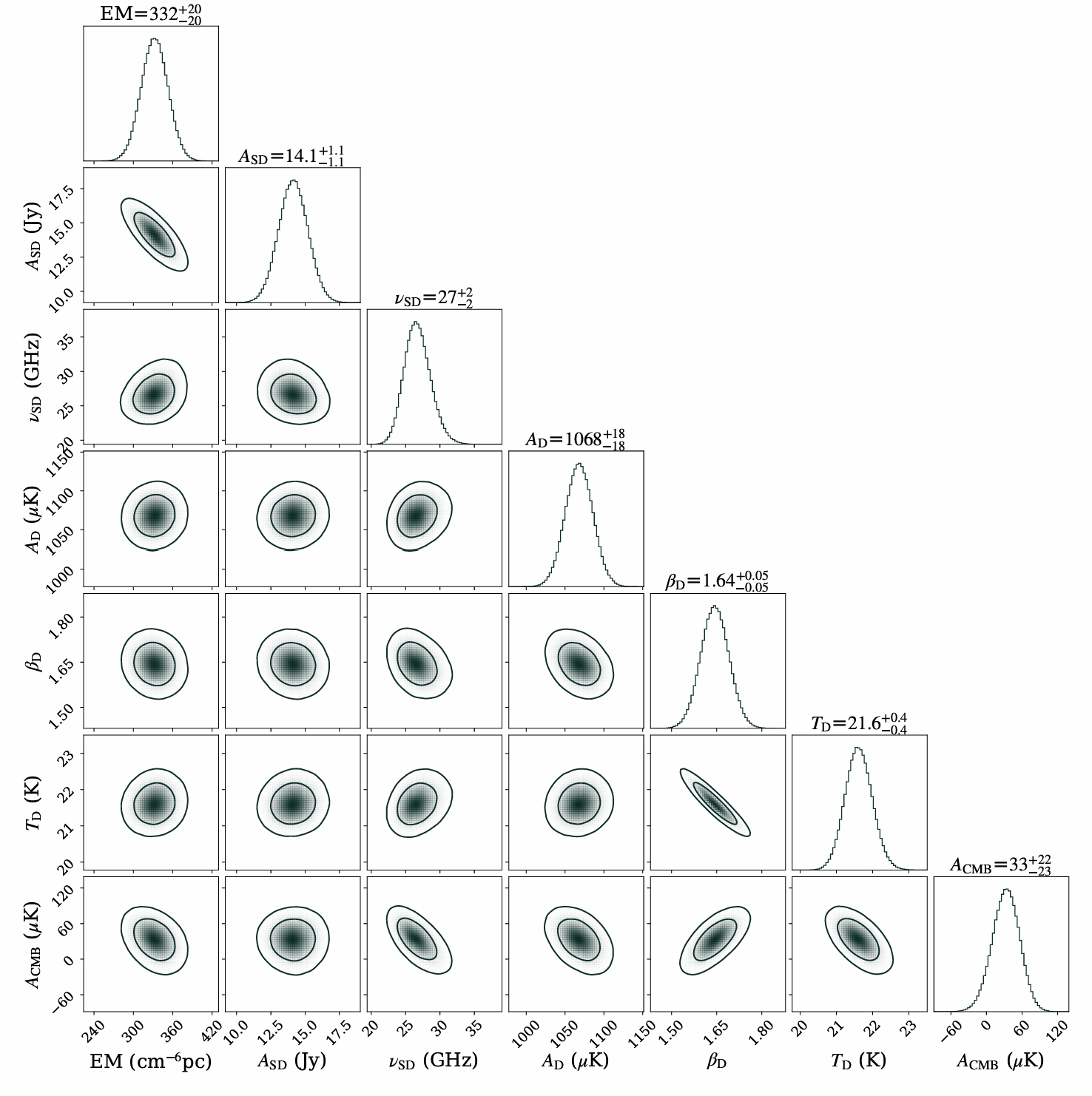}
    \caption{Posterior distribution for the SED fit using a spinning dust component (equation \ref{eq:spinning-dust-model}), with a fixed width of 0.7 (dimensionless). The posteriors are generally well-constrained and approximately Gaussian.  As expected, we observe strong degeneracy between the dust temperature, $T_\text{D}$ and dust spectral index $\beta_\text{D}$. A weaker degeneracy is also present between the spinning dust amplitude, $A_\text{SD}$ and the free-free emission measure (EM), which could be improved by more low-frequency data.}
    \label{fig:corner-plots-sd}
\end{figure*}

\begin{figure*}
    \includegraphics[width=0.85\textwidth]{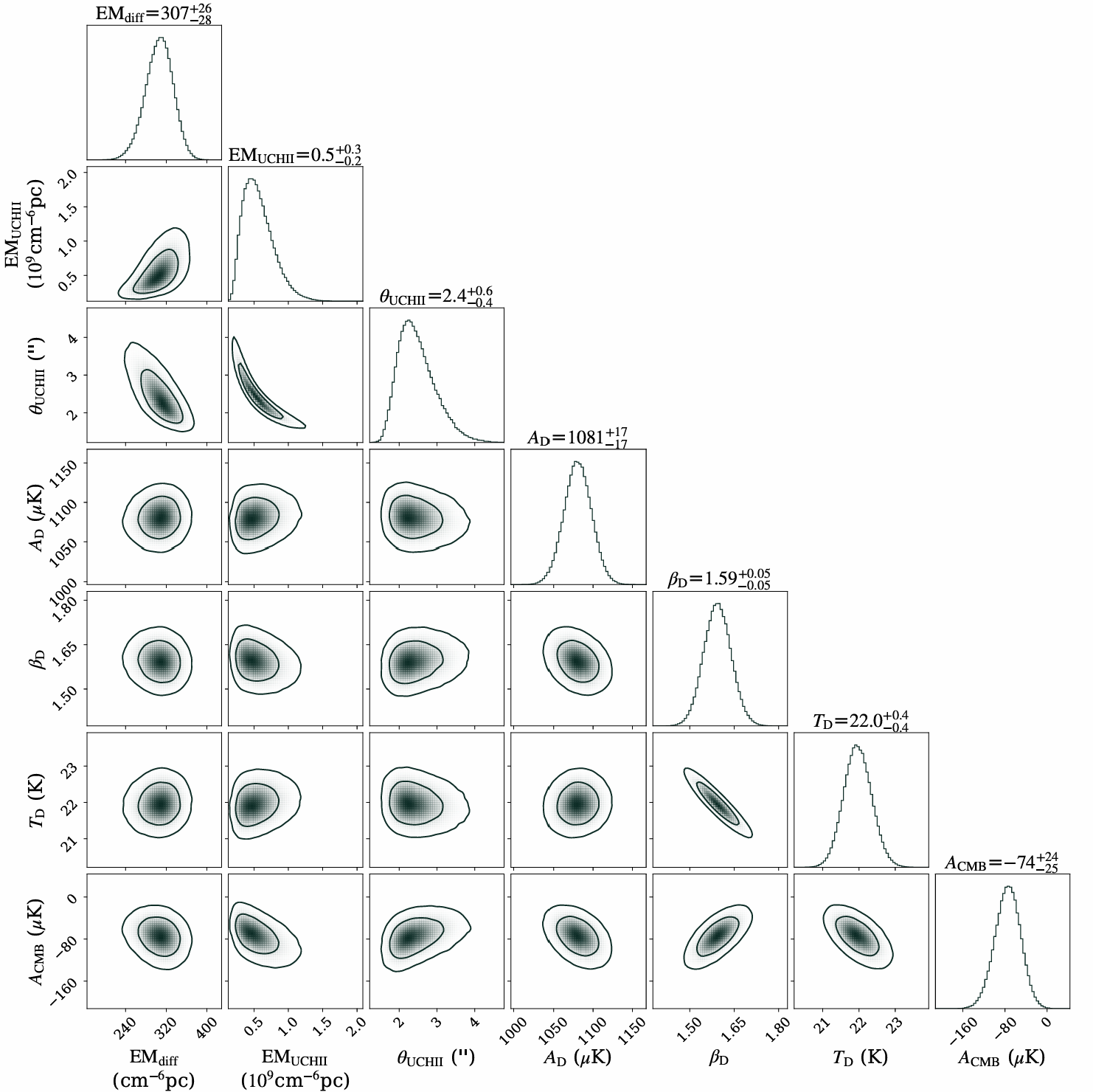}
    \caption{Posterior distribution for the SED fit using an optically thick free-free component (equation \ref{eq:uchii-model}). Compared to the model with a  spinning dust component, these posteriors are less well-constrained, more non-Gaussian, and exhibit stronger parameter degeneracies.}
    \label{fig:corner-plots-uchii}
\end{figure*}

\begin{table*}
    \centering
    \caption{MCMC Sampling results. For each model, we report the median parameters and errors. We also include the calculated WAIC and $p_\text{WAIC}$; we conclude that the spinning dust model is preferred based on its lower WAIC and fewer effective parameters, $p_\text{WAIC}$.}
    \label{tab:best-fit-params}
    \include{tables/tab5_mcmcSamplingResults}
\end{table*}

We use least-squares fitting and Markov Chain Monte Carlo (MCMC) analysis to determine our best-fit parameters and evaluate the posterior distribution of each set of emission models. 
The SED fits are shown in Figure \ref{fig:seds-and-fits} and the MCMC posteriors are shown in figures \ref{fig:corner-plots-sd} and \ref{fig:corner-plots-uchii}. 
We report the median values and 1$\sigma$ errors from the MCMC samples in Table \ref{tab:best-fit-params}.
We compare the model with a spinning dust component (Eq. \ref{eq:spinning-dust-model}) against the model with an ultra-compact H\,\textsc{ii} component (Eq. \ref{eq:uchii-model}). We evaluate the relative quality of the two models using the ``Widely Applicable'' Watanabe-Akaike Information Criterion (WAIC) \citep{watanabe_asymptotic_2010}.
The WAIC is a fully Bayesian information criterion used to estiamte and compare the predictive accuracy of models, and includes a penalty term for model complexity. 
It is not the actual WAIC value that is important, but rather the difference $\Delta\text{WAIC}$ between the models we wish to compare, with a lower value indicating a better model.

The model with a spinning dust component has a WAIC of 38, and the model with a UC H\,\textsc{ii} component has a WAIC of 47. A lower value is considered better and our difference of 9 significantly favors the spinning dust scenario. 
We also compare $p_\text{WAIC}$, which can be interpreted as an approximation of the effective number of unconstrained parameters in the model.
The models using only spinning dust and only UCH\,\textsc{ii} have a $p_\text{WAIC}$ of 8 and 12, respectively.
It is unsurprising that $p_\text{WAIC}$ exceeds the number of free parameters in both models, considering the known degeneracy between $T_\text{D}$, $\beta_\text{D}$, and $A_\text{CMB}$, as well as additional degeneracy between EM$_\text{UCHII}$ and $\theta_\text{UCHII}$ in the latter model. 
Therefore, we consider the model with a spinning dust component superior to the model with a UCH\,\textsc{II} component based on its higher WAIC, lower $p_\text{WAIC}$, and less degeneracy in the posterior distribution.
For the model with a spinning dust component, we fit a spinning dust amplitude of $14.1\pm1.1$~Jy and a peak frequency of $27\pm2$~GHz.
For the model with a UC H\,\textsc{ii} component, we fit an EM of $0.5^{+0.3}_{-0.2}\times 10^9$ cm$^{-6}$~pc and $\theta_\text{UCHII}$ of $2.4^{+0.6}_{-0.4}$ arcsec.
At a distance of 910 pc, $\theta=2.4$~arcsec corresponds to 0.01 pc, so these parameters are consistent with hyper-compact H\,\textsc{ii} emission.

\section{Morphological analysis}\label{sec:morphological-analysis}
We use our arcminute resolution GBT data to further investigate the emission mechanisms in the region using similarly high-resolution multi-frequency maps that trace gas and dust emission. 
Our objective is to locate excess 13 GHz emission, then check for spatial coincidence with emission by or abundance of small grains that could produce spinning dust emission.

\subsection{Multi-Wavelength Maps and excess 13 GHz emission}
\label{ssec:morphological-comparison}

\begin{figure}
    \centering
    \includegraphics[width=\columnwidth]{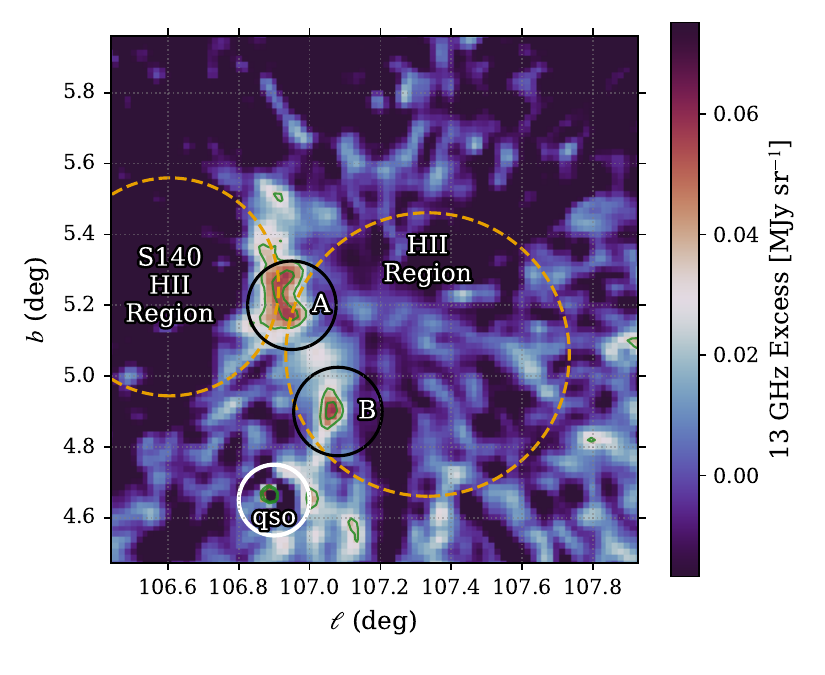}
    \caption{Ku-band excess map, constructed by subtracting the 4.5~GHz GBT C-band map from the 13.1~GHz GBT Ku-band map. The map radius is 45 arcmin, matching the photometry aperture illustrated in Figure \ref{fig:kumap-with-noise}. Both maps were background-subtracted and the Ku-band map was smoothed to C-band resolution. The green countours represent the $2\sigma$ and $3\sigma$ levels, defined with respect to the background RMS level in the difference map. There are three regions with surface brightness $\geq3\sigma$ significance. Source ``qso'' in the lower left corner is the quasar LQAC 335+062 001. We interpret the other two sources at G106.95+5.2 (region A) and G107.08+4.9 (Region B) as potential sources of spinning dust emission.}
    \label{fig:ku-band-excess}
\end{figure}
\begin{figure*}
    \centering
    \includegraphics[width=\textwidth]{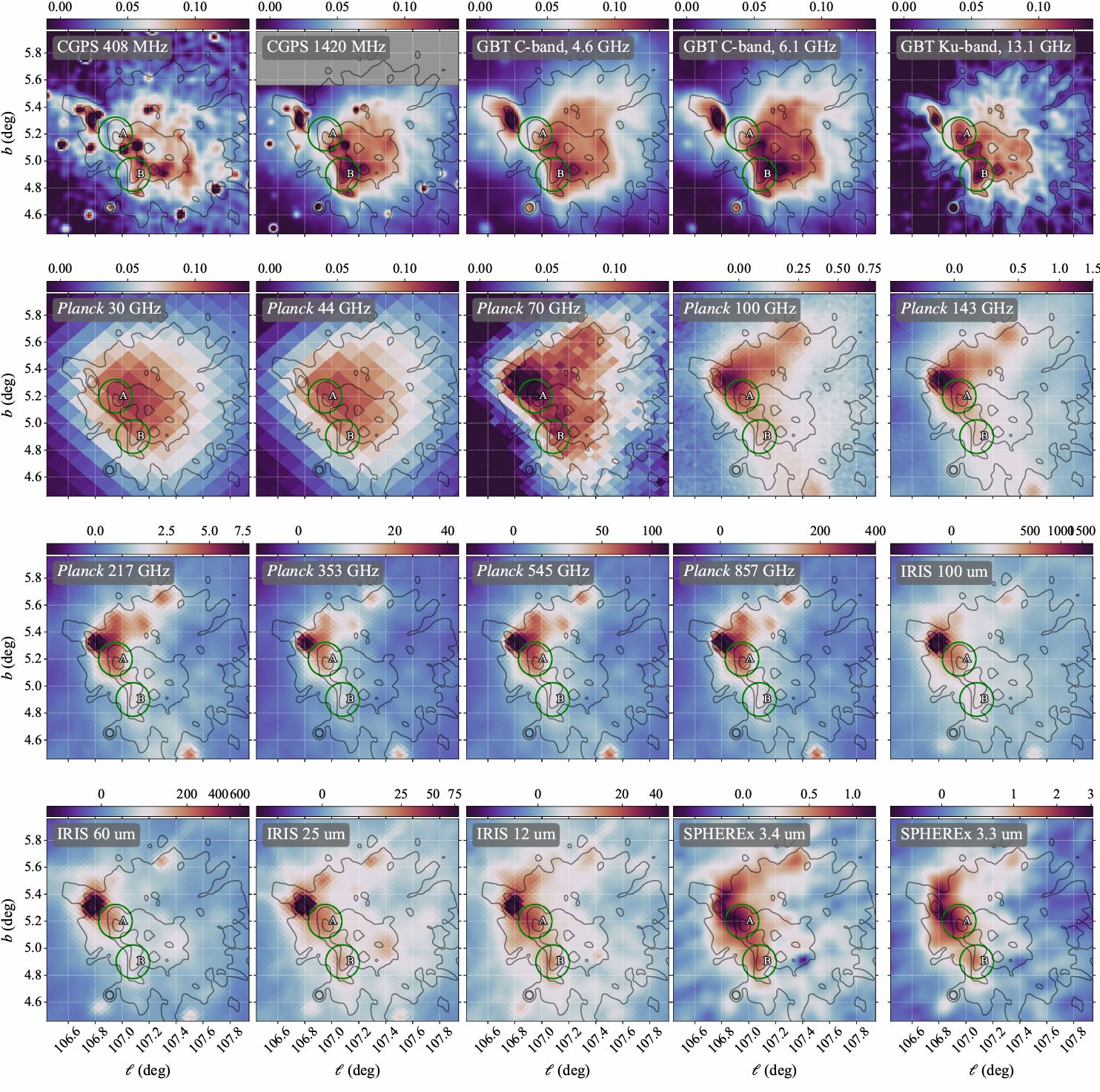}
    \caption{Maps of the G107.2+5.20 region with a 45 arcmin radius with contours from the Ku-band map overlaid.
    Region A and B from Figure \ref{fig:ku-band-excess} are overlaid.
    We expect a spinning dust signal to be negligible at 4--6 GHz, observable at 13 GHz, and spatially correlated with dust emission.
    The CGPS maps trace synchrotron and diffuse free-free emission. While the CGPS 1420 MHz map does not fully cover our region, it covers the primary regions of interest so we include it anyway.
    The C-band (5 GHz) map is dominated by free-free emission.
    The primary features are the bright H\,\textsc{ii} region S140 at G106.8+5.3 and the large central cloud of diffuse H\,\textsc{ii}. 
    The Ku-band (12 GHz) map is a mix of free-free and AME. 
    S140 and the H\,\textsc{ii} cloud are both visible.
    The \textit{Planck} 30 GHz map includes AME and free-free emission, though the resolution is poor. 
    The \textit{Planck} 353 GHz map traces thermal dust emission. 
    The IRIS 100~$\mu$m, 25~$\mu$m, and 12~$\mu$m maps trace emission from large, medium, and small dust grains, respectively.
    While S140 dominates the dust emission in all maps, region A is also coincident with relatively strong dust emisson. 
    Region B shows weak thermal dust emission, but does coincide with emission in the IRIS and SPHEREx maps.
    We conclude based on the presence of dust emission that region A and B are consistent with spinning dust emission; however, the strongest dust emission does not trace the Ku-band excess.
    The ``radio point source'' around G106.9+4.65 has one matching radio source in NED and appears in the Large Quasar Astrometric Catalog (LQAC-03), LQAC 335+062 001. 
    }
    \label{fig:morphological-comparison}
\end{figure*}

We consider the possibility that the spinning dust emission is not uniformly diffuse across our region, but could appear as localized excess 13 GHz emission relative to the 4-6 GHz emission maps (Fig. \ref{fig:ku-band-excess}). 
If this is the case, we would also expect the emission to be spatially correlated with small dust grains that could produce spinning dust emission.

Fig. \ref{fig:morphological-comparison} shows multi-wavelength maps at $\sim4$ arcmin resolution (or native resolution, if greater than 4'), with contours from the Ku-band maps overlaid.
The CGPS and GBT C-band maps are predominantly free-free emission, while the GBT Ku-band map includes free-free plus spinning dust.
Using a peak-finding algorithm, we detect one emission peak, G106.95+5.19, in the Ku-band map with no C-band counterpart that is spatially coincident with emission from the \textit{Planck} 353 GHz, IRIS 12 $\mu$m, and our SPHEREx 3.3 and 3.4~$\mu$m emission maps, which trace thermal dust, small grains, and the smallest PAHs, respectively. 

We also construct a ``difference map'' (Figure \ref{fig:ku-band-excess}) to directly visualize regions of excess 13 GHz emission relative to 4 GHz emission. 
Our GBT 4 GHz maps are free-free dominated, and our GBT 13 GHz maps also contain substantial free-free emission. 
Since spinning dust has a rising 4--13 GHz spectra, we expect excess 13 GHz emission relative to 4 GHz.
We begin by smoothing both maps to a common 4 arcmin resolution and regridding to a common pixel scale, then background-subtracting each map using the median flux value in the photometry annulus as described in Section \ref{ssec:aperture-photometry}. 
We apply 2 and 3 $\sigma$ contours based on the standard deviation measured in a background annulus with radius 55-75' on the 13 GHz excess map.

We detect three excesses at $\geq 3 \sigma$ significance.
The strongest excess, located at G106.9+4.66 is a known quasar, LQAC 335+062~001. 
The next excess, ``region A'' located at G106.95+5.19, coincides with the aforementioned dust-correlated source in the GBT 13 GHz intensity map.
Region A overlaps the edges of two HII regions identified in the WISE catalogue: S140 and the large, central HII region in the map at G107.333,+5.061, both outlined in Figure \ref{fig:ku-band-excess} \citep{anderson_wise_2014}. 
The final excess, ``region B'', is located at G107.08+4.9. 
Region B is located near the edge of the large central HII region.
This source is not correlated with the brightest dust emission in any map--they are all dominated by S140.
However, it does coincide with emission in the IRIS 12 $\mu$m and SPHEREx 3.4 and 3.3 $\mu$m maps, which trace small grains and the smallest PAHs, respectively.
Therefore, we consider region A and region B to be candidate sources of spinning dust emission.

\subsection{Tracing compact H\,\textsc{ii}: H~$\alpha$ and 21 cm continuum maps}
\label{ssec:compact-HII-test}

\begin{figure*}
    \centering
    \includegraphics[width=0.8\textwidth]{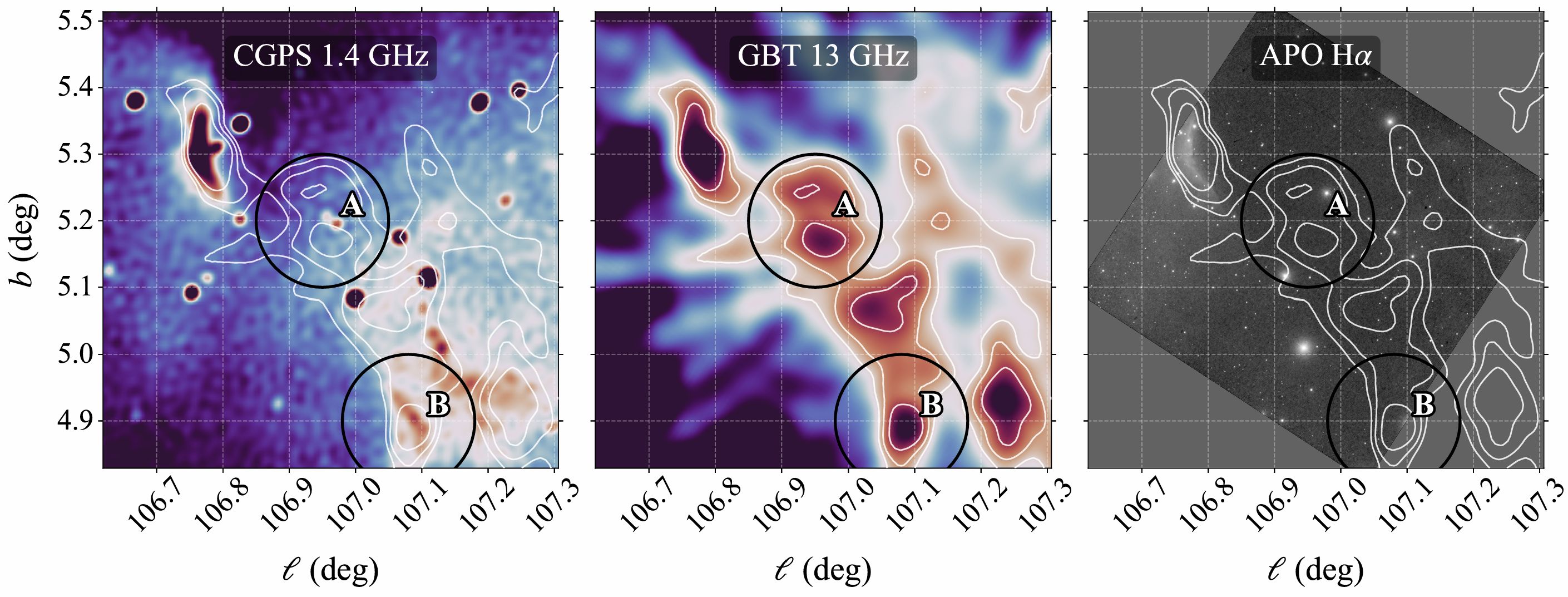}
    \caption{Comparison of H~$\alpha$ observations with GBT Ku-band and CGPS 1420 MHz maps, showing no significant signal in region A in H~$\alpha$. Overlayed contours are the 70th, 85th, and 95th percentiles of the GBT data. 
    The strong S140 region is visible in the upper left of the image, and regions A and B, defined in Fig. \ref{fig:ku-band-excess} are circled in black.
    The source in region A at $\sim106.95+5.2$ in the CGPS 1.4 GHz map is NVSS J222112+631856. The SPECFIND tool on the Vizier database reports a 1.4-9 GHz SED with a spectral index of $\alpha=-0.7$ \citep{stein_specfind_2021}, so we consider it unlikely to be the source of the 13 GHz excess emission.
    Region B was not fully covered by our H-alpha observations, but the CGPS data reveal compact 1.4 GHz emission which could be consistent with optically thick free-free emission.
    }
    \label{fig:ha-comparison}
\end{figure*}

\begin{figure*}
    \includegraphics[width=\textwidth,trim={0cm 2.2cm 0cm 0cm}, clip]{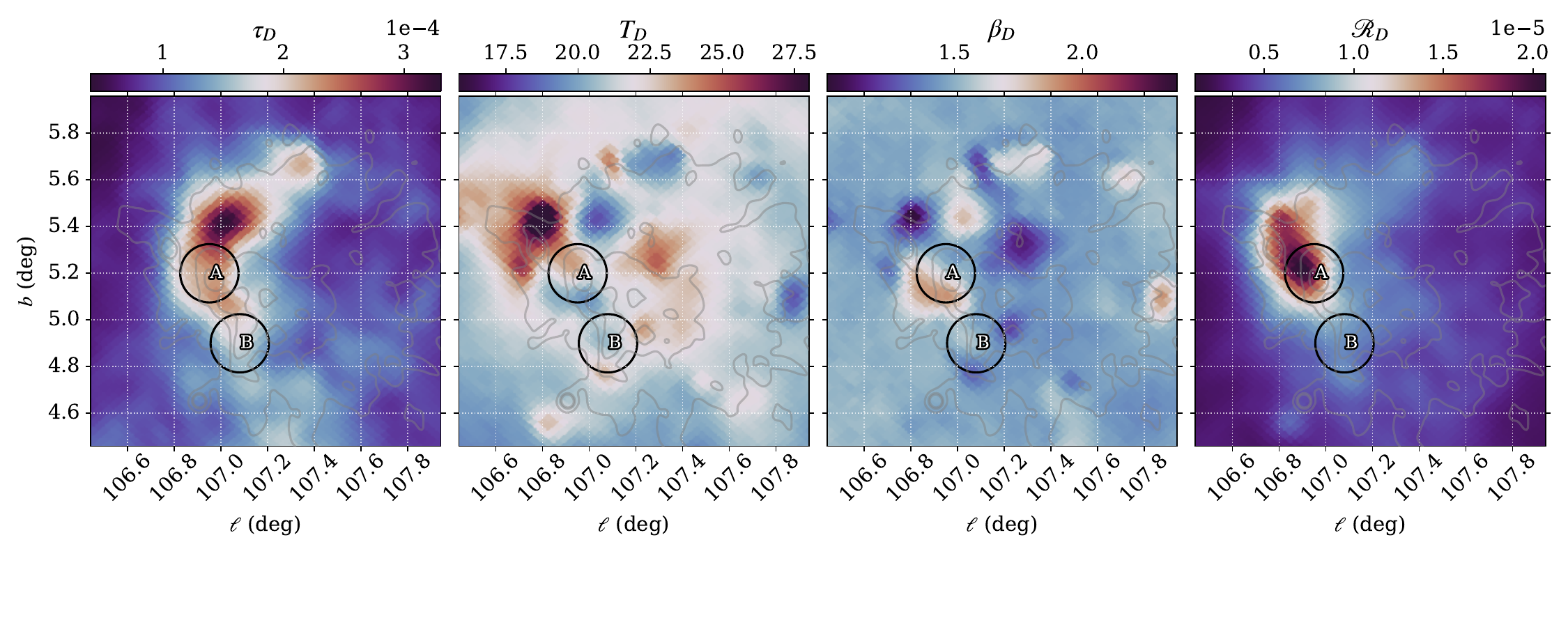}
    \caption{\textit{Planck} GNILC-derived dust parameter maps. The opacity map largely traces the L1204 dark cloud. The ``hot spot'' in the temperature map coincides with S140, as expected. The brightest region in the dust radiance map overlaps with region A.
    }
    \label{fig:planck-dust-params}
\end{figure*}

\begin{figure*}
    \centering
    \includegraphics[width=.9\textwidth]{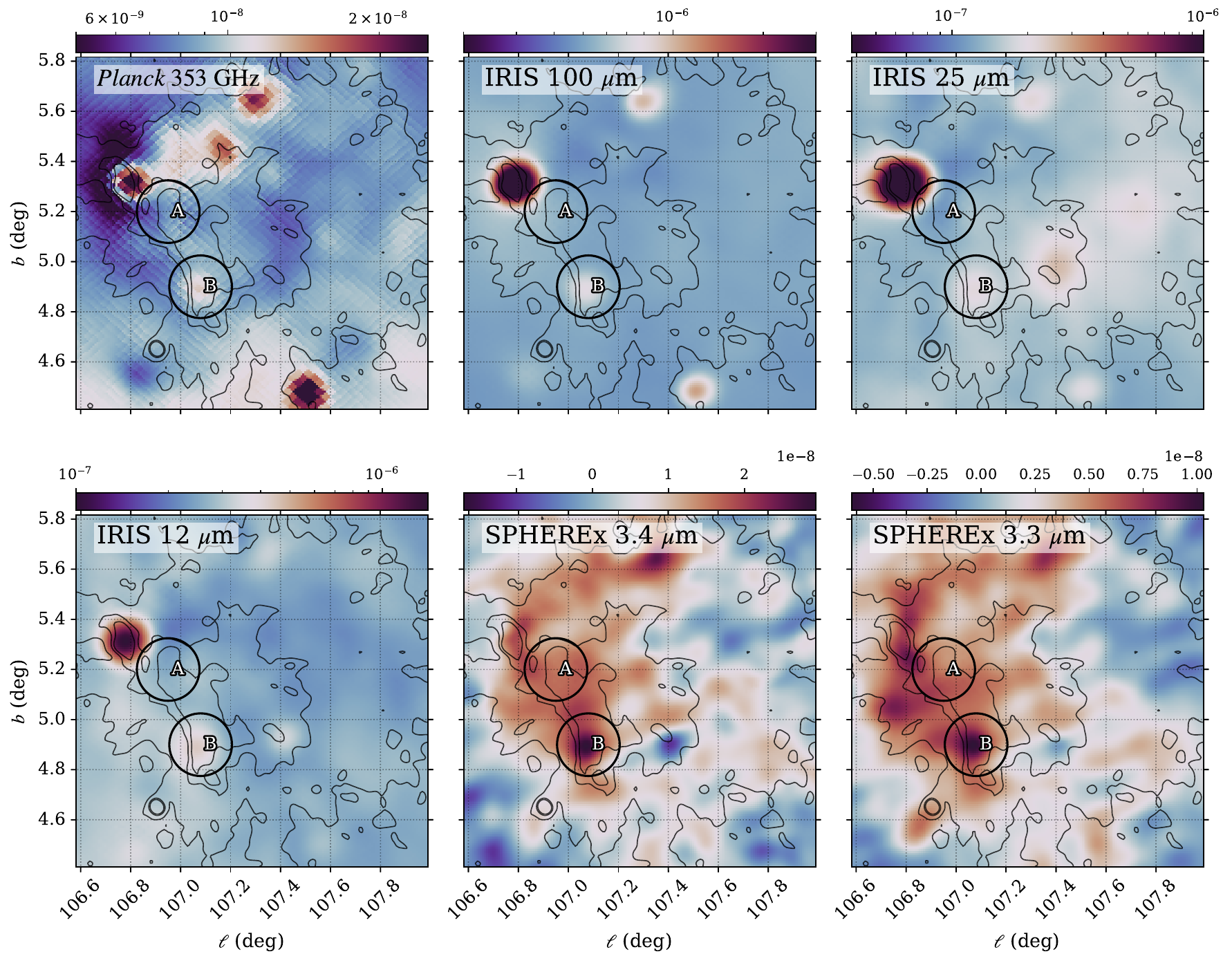}
    \caption{Abundance maps derived from the \textit{Planck} dust radiance map with the contours from the Ku-band map (\ref{fig:kumap-with-noise}) overlaid. 
    The regions of 13 GHz excess from Fig. \ref{fig:ku-band-excess} are circled in black.
    \textit{Planck} 353 GHz traces thermal dust emission; IRIS 100, 25, and 12 $\mu$m trace large--, intermediate--, and small--sized grains, respectively. 
    The IRIS 12 $\mu$m map traces small grains and the bandwidth includes some PAH features.
    The SPHEREx 3.3 and 3.4 $\mu$m maps were constructed to cover the 3.3 $\mu$m aromatic and 3.4 $\mu$m aliphatic PAH emission features, respectively, so these maps estimate the abundance of the smallest PAHs.
    While the abundance maps are not globally correlated with the Ku-band excess map, we note that region B coincides with peak 3.3 and 3.4 $\mu$m abundance. 
    }
    \label{fig:abundance-maps}
\end{figure*}

While our SED analysis favors spinning dust over HC-H\,\textsc{ii} as the global emission mechanism, we wanted to check whether optically-thick free-free emission may be contributing locally to our source at G106.95+5.19 (region A).
\cite{abitbol_constraining_2018} suggested that high-resolution H~$\alpha$ observations of this region could be used to trace free–free emission and therefore help disentangle the two emission mechanisms being considered.
H~$\alpha$ emission arises from recombination in photoionized hydrogen and is therefore closely related to the emission measure that also governs free–free radiation at radio frequencies \citep{dickinson_towards_2003, finkbeiner_full-sky_2003}.
In 2018, only coarse resolution data was available, so no analysis was done in \cite{abitbol_constraining_2018}.
For this study, we chose to test the Abitbol hypothesis by making  narrow-band H~$\alpha$ and continuum observations of the G106.95+5.19 region with the 0.5-m Astrophysical Research Consortium Small Aperture Telescope (ARCSAT) at Apache Point Observatory.
In October 2025, we collected approximately two hours of data.
Approximately half of the data was collected using a BYU H$\alpha$ emission line filter with a 3~nm pass band centered on the 656.3 nm.
The remaining observations used a continuum filter, which transmits wavelengths near 645~nm.
The rightmost panel in Figure~\ref{fig:ha-comparison} shows the continuum-subtracted H~$\alpha$ image constructed from calibrated and stacked exposures.
We found no significant H~$\alpha$ excess in region A, and the continuum-subtracted image shows no coherent H~$\alpha$ structure spatially coincident with the Ku-band excess.
Given the clear detection of strong H~$\alpha$ emission from the nearby S140 region, the lack of H~$\alpha$ emission near G106.95+5.19 shows that any compact hydrogen emission source is either absent or strongly dust obscured (which is likely for ultra- and hyper-compact free-free sources).

Additionally, Figure \ref{fig:ha-comparison} shows the CGPS 1.4 GHz map at native resolution, which can also be used as an H\,\textsc{ii} tracer.
In the 1.4~GHz map, there is one point source that is roughly coincident with region A.
Using the NASA Extragalactic Database (NED), we identified it as NVSS J222112+631856, and the SPECFIND tool on the Vizier database reports a 1.4--9 GHz SED with a spectral index of $\alpha=-0.7$ \citep{stein_specfind_2021}.
A hyper-compact H\,\textsc{ii} source should have a rising 1--9 GHz SED, so we consider it unlikely that NVSS J222112+631856 is the source of HC H\,\textsc{ii} emission contributing to a Ku-band excess in region A.

\subsection{Ku-band excess versus intensity, dust radiance, and small-grain abundance} \label{ssec:planck-dust-params}

\cite{hensley_case_2016}
used degree-scale maps of AME and dust parameters derived from \textit{Planck} data, and found that while AME is spatially correlated with dust tracers in general, it is best correlated with dust radiance.
Additionally, to probe the particular grain population(s) that may give rise to spinning dust emission, \cite{hensley_case_2016} normalized the intensity maps for their dust tracers by the dust radiance to find the ``AME efficiency'' and found that PAH fraction did not trace AME especially well.
\cite{chuss_tracing_2022} perform a similar analysis in the $\lambda$-Orionis region, focusing on 3.3$\mu$m PAH emission--the shortest wavelength of the known PAH features corresponding to the smallest PAHs. 
They found that AME correlated better with far-infrared dust tracers than with PAH emission, and and suggest that this difference may arise from the distinct sensitivity of rotational (AME) and vibrational (IR) emission mechanisms to the local radiation environment.
The angular resolution of these studies were limited to $\geq$ 0.5$^\circ$ by the \textit{Planck} LFI and DIRBE maps.

Using our 13 GHz Ku-band excess map, we conduct a similar, qualitative ``case study'' at higher resolution. 
We use the Generalized Needlet Internal Linear Combination (GNILC)-derived \textit{Planck} dust parameter maps at native $\sim4$ arcmin resolution \citep{aghanim_planck_2016}.
Figure \ref{fig:planck-dust-params} shows the four dust parameter maps: opacity $\tau_\text{D}$, temperature $T_\text{D}$, spectral index $\beta_\text{D}$, and dust radiance $\mathscr{R}_\text{D}$.
Rather than estimating PAH fraction from wideband data, we construct mosaiacs from publicly available SPHEREx data from ``quick release'' (QR) 2 \citep{spherex_team_spherex_2025}. 
The SPHEREx survey is an all sky survey covering the near-infrared in 102 spectral channels using linear variable filters with an angular resolution of 6.2 arcsec. The mission launched in spring of 2025 and will collect data for two years, mapping the sky four times. While there are six spectrometers (bands), for the purposes of this study, we make use of the fourth band, covering 2.42–3.82 $\mu$m with resolving power of 35. 
Following methods described in \cite{cukierman_spectral_2026} and \cite{murgia_spherex_2026}, we compile SPHEREx cutouts in which both the physical coordinates and wavelength overlap our desired range. We apply a contour filter to remove zodiacal light from each cutout. To create the mosaics, we divide our wavelength range into 8 equal bandwidth sub-channels and re-project to a spatial grid for each sub-channel, essentially creating a three dimensional spectral data cube. For spatial pixels that are covered in multiple sub channels, we interpolate between them to central wavelength and project onto the final map. 
The 3.4 and 3.3 $\mu$m maps have 80.6 per cent and 90.7 per cent coverage, respectively, and we interpolate over missing pixels.
Our continuum-subtracted 3.4 and 3.3 $\mu$m emission maps are shown in the last two panels of Figure \ref{fig:morphological-comparison}. 
The continuum is calculated by a linear fit between $3.15 \pm 0.05$ $\mu$m and $3.60\pm 0.05$ $\mu$m, and then subtracted from the $3.3 \pm 0.05$ $\mu$m and the $3.4 \pm 0.03 \mu$m maps.
The tolerances for each map are chosen to include only the relevant PAH emission feature while maximizing the number of contributing pixels. 
In Figure \ref{fig:abundance-maps}, we present 3.4 and 3.3 $\mu$m PAH abundance maps, alongside other dust abundance maps.
The abundance maps were calculated by dividing the intensity maps by the GNILC-derived \textit{Planck} dust radiance map.

Neither region A nor region B coincides with any peak dust emission (S140 dominates the intensity maps); still, both regions coincide with some dust emission. 
Region A coincides with fairly bright thermal dust, IRIS 12 $\mu$m, and SPHEREx 3.4 and 3.3 $\mu$m.
Region B is less bright overall, but coincides with IRIS 12 $\mu$m and SPHEREx 3.4 and 3.3 $\mu$m emission.
Region A coincides with the peak of the \textit{Planck} dust radiance map. Region A does not correspond to any peaks in our abundance maps. 
In contrast, region B occupies a region of relatively low dust radiance, yet coincides with the peak of our 3.3 and 3.4 $\mu$m PAH abundance maps.

\section{Discussion}\label{sec:discussion}

We fit two models to the 408 MHz--3THz SED which could explain the observed 30 GHz excess, using either a spinning dust or ultra-compact H\,\textsc{ii} emission component. 
Both models have similar residuals, though the Ku-band data is more consistent with the spinning dust model. The UCH\,\textsc{ii} model has broader and more asymmetric posterior distributions.
For the spinning dust and UCH\,\textsc{ii} models, we measure a WAIC of 38 and 47, respectively, and a $p_\text{WAIC}$ of 8 and 12, respectively, with lower values indicating that the spinning dust model is better.
For the spinning dust model, we fixed the width to 0.7 and the MCMC sampling finds an amplitude of $14.1\pm1.1$~Jy and a peak frequency of $27\pm2$~GHz.
With the addition of our Ku-band data, we constrain the rising edge of the SED and find a lower peak frequency than the $30.9\pm1.4$ GHz reported by \cite{abitbol_constraining_2018}. \cite{poidevin_quijote_2023} report a peak frequency of $25.6\pm2.7$ GHz and an amplitude of $18.5\pm0.7$ Jy. Our peak frequency is consistent. The amplitude discrepancy may be due to differences in our free-free EM level ($330\pm20$ versus $217\pm6$), which could be better constrained with more low-frequency data.
It is possible that the region contains a combination of UC H\,\textsc{ii} and spinning dust emission, since there is known to be optically thick free-free emission in the region, e.g. embedded in S140. 
We considered a five-component model with a combination of spinning dust and optically thick free-free, but the spinning dust component remained similar to the SD-only model while the UC H\,\textsc{ii} parameters are poorly constrained by our low-frequency data, so we did not pursue it further. 
There are not many known UC H\,\textsc{ii} regions within this region \cite{anderson_wise_2014}, so although the UC H\,\textsc{ii} region might dominate the small sub region A, it is unlikely that UC H\,\textsc{ii} will be the main contributor to the emission when integrating over an aperture of radius 45 arcmin.

In our morphological analysis of the maps at $\sim4$ arcmin resolution, we observe one compact source of emission in the Ku-band maps that has no C-band analog.
It is spatially correlated with thermal and IR dust emission, consistent with the spinning dust scenario.
Using our ``difference maps'' as a proxy to trace AME, we find that this source, G106.95+5.19 (``region A''), is one of two sources of excess 13 GHz emission consistent with spinning dust that have $\geq3\sigma$ significance in peak surface brightness.
Region A is spatially coincident with the peak dust radiance observed in the \textit{Planck} dust parameter maps (Figure \ref{fig:planck-dust-params}).
The second source, G107.08+4.90 (``region B''), is spatially coincident with the peak PAH abundance estimated from our SPHEREx 3.3 and 3.4 $\mu$m maps. 
If region A and B are indeed sources of spinning dust emission, our region demonstrates a resolved example of how correlation with dust tracers may vary between environments.
We leave a detailed correlation analysis between the 13 GHz excess emission and PAH emission or abundance to future work.

However, it is possible that UCH\,\textsc{ii} or background objects contribute some of the Ku-band excess. 
We see no evidence of compact emission in our H~$\alpha$ observations, however this is inconclusive since the dark cloud could absorb emission. 
We see one radio source in the CGPS 1.4 GHz map; however, this is identified as NVSS J222112+631856 and has a measured 1.4-9 GHz spectral index of $\alpha=-0.7$, so we consider it unlikely to be the source of the excess 13 GHz emission.

Because our observing strategy was designed to do both matched-resolution aperture photometry with CMB data sets and direct imaging of 13 GHz excess, we required large maps and thus balanced sensitivity against observing time (which scales with map radius squared). 
Now that two sources of significant Ku-band excess have been identified with different dust populations, it would be interesting to do follow-up observations of these sub-regions to study the relationship in more detail. 
In particular, the CO Mapping Array Project (COMAP) would be well-suited to aid in this effort, with 5' resolution at 30 GHz, enabling photometry on much smaller regions than the \textit{Planck} data allows \citep{cleary_comap_2022,lamb_comap_2022,rennie_comap_2022}.
And, because COMAP can sample both the rising and falling edge of the AME peak, it would be better suited to estimate limits on contributions from optically thick free-free contribution, which has a much flatter SED above 30 GHz than spinning dust.

AME is generally referred to as diffuse; however regions A and B are more compact sources than the CMB data sets can resolve. 
It would be interesting to see whether this structure is peculiar to our region, or typical across a sample of sources. 
Forthcoming observations from facilities such as COMAP and the Simons Observatory Large Aperture Telescope \citep{zhu_simons_2021}, with $\leq10$~arcmin resolution near 30 GHz, will likely offer new and exciting insights on this front for sources in their declination range, and may benefit from similar C- and Ku-band observations as the ones presented here to constrain the low-frequency edge of the AME spectrum. 
With higher resolution near 30 GHz, the follow-up 4--12 GHz maps could be smaller and more efficiently obtained with the GBT.

\section{Conclusions}\label{sec:conclusions}
Globally, our region is best-fit by a spinning dust model with an amplitude of $14.1\pm1.1$ Jy and a peak frequency of $27\pm2$. 
Using our high-resolution maps, we identify two sources of significant excess 13 GHz emission consistent with spinning dust: G106.95+5.19 (``region A'') and G107.08+4.9 (``region B'').
Region A is near the edge of the L1204 dark cloud and the S140 H\,\textsc{ii} region.
Using the \textit{Planck} dust parameter maps, we see that region A coincides with the dust radiance peak, but where dust temperature is moderate-low (22~K) and opacity is relatively high compared to the rest of the region.
This is consistent with the \cite{hensley_case_2016} finding that dust radiance is a superior tracer of AME on degree scales, and consistent with evidence that spinning dust is associated with transitional/interface regions (e.g. \cite{casassus_resolved_2021} and references therein). 
In contrast, region B is associated not with peak dust radiance, but rather with peak PAH abundance estimated from 3.3 $\mu$m SPHEREx maps. 
This is consistent with PAHs as an AME ``carrier particle.''
There remains some ambiguity as to the contribution of optically thick free-free emission. The aim of this study was to measure the SED at 13 GHz and identify locations of excess Ku-band emission. We leave a detailed analysis of this to future work, and suggest that COMAP data sampling the falling edge of the SED would help disentangle the emission mechanisms.

If both region A and B are in fact spinning dust emission in two different environments, the G107.2+5.20 region could be an example of why low-resolution all-sky studies struggle to identify a single, best tracer of AME.
Furthermore, this more compact structure we see could have important implications for CMB foregrounds. 
Spinning dust is generally treated as unpolarized, and $1\%$ limits on the polarization fraction have been observed over large ($\sim0.5^{\circ}$) apertures. 
However, if spinning dust is polarized locally but incoherent globally, this unresolved structure could bias CMB component separation.
This work underscores the need for further high-resolution observations of AME regions that cover a diverse range of environments and potential carrier particles.

\section*{Acknowledgements}

We are grateful to Larry Morgan for support in developing the GBT observing plan. We also thank the GBT operators, especially Brenne Gregory for her assistance with data volume management.
We thank Brandon Hensley for helpful conversations and advice on analysis and modeling.
We also thank Ilse Cleeves for helpful conversations and suggestions over the course of this project.
Thank you to Jimmy Davidson for support in developing the APO observing plan, and the APO staff for their support during observations.
Support for this work was provided to J.E.S. by the NSF through the Grote Reber Fellowship Program administered by Associated Universities, Inc./National Radio Astronomy Observatory, as well as the P.E.O. scholar award.
The National Radio Astronomy Observatory and Green Bank Observatory are facilities of the U.S. National Science Foundation operated under cooperative agreement by Associated Universities, Inc.
Support for this work was provided to S.E.H. from UKSA (LiteBIRD UK ST/Y005945/1).

Facilities:
We acknowledge the use of data provided by the Centre d'Analyse de Données Etendues (CADE), a service of IRAP-UPS/CNRS (http://cade.irap.omp.eu, \cite{paradis_dark_2012}).
The research presented in this paper has used data from the Canadian Galactic Plane Survey, a Canadian project with international partners, supported by the Natural Sciences and Engineering Research Council.
Some of the presented results are based on observations obtained with the QUIJOTE experiment, with doi.org/10.26698/quijote-mfi-dr1. 
This work is based on observations obtained with Planck (http://www.esa.int/Planck), an ESA science mission with instruments and contributions directly funded by ESA Member States, NASA, and Canada. 
This research has made use of data obtained from the Diffuse Infrared Background Experiment (DIRBE) developed by NASA's Goddard Space Flight Center, and archived at the Legacy Archive for Microwave Background Data Analysis (LAMBDA) and the Infrared Science Archive (IRSA).
This publication makes use of data products from the Spectro-Photometer for the History of the Universe, Epoch of Reionization and Ices Explorer (SPHEREx), which is a joint project of the Jet Propulsion Laboratory and the California Institute of Technology, and is funded by the National Aeronautics and Space Administration.
Some of the results are based on observations obtained with Apache Point Observatory's 0.5-m Astrophysical Research Consortium Small Aperture Telescope.

AI tools were used on a limited basis for literature search cross-checking and routine technical tasks such as code and plot formatting. The analysis design, modeling, interpretation, and writing were conducted by the authors.

\section*{Data Availability}

The raw data generated for this article is publicly available on the GBT archive under project code AGBT21B\_241. The maps produced for this article can be shared on reasonable request to the corresponding author.
The H~$\alpha$ observations can be shared upon reasonable request to the corresponding author. 
All other data are publicly available.


\bibliographystyle{mnras}
\bibliography{references} 

\bsp	
\label{lastpage}
\end{document}

%% file: tables/tab1a_spwDefinitions_kuband.tex
\begin{tabular}{c|c|c|c|c}
\hline
SPW & $\nu_c$ & $\Delta\nu_r$ & $\Delta\nu_s$ & FWHM$_c$ \\
     & (GHz) & (GHz) & (GHz) & (arcmin) \\
\hline
    A &	12.10 & 11.62--13.12 &	11.75--12.45 &	1.04 \\
    B &	13.10 & 12.38--13.87 &	12.75--13.45 &	0.96 \\
    C &	13.85 & 13.12--14.62 &	13.50--14.20 &	0.91 \\
    D &	14.60 & 13.88--15.37 &	14.25--14.95 &	0.86 \\
\hline
\end{tabular}

%% file: tables/tab1b_spwDefinitions_cband.tex
\begin{tabular}{c|c|c|c|c}
\hline
SPW & $\nu_c$ & $\Delta\nu_r$ & $\Delta\nu_s$ & FWHM$_c$ \\
     & (GHz) & (GHz) & (GHz) & (arcmin) \\
\hline
    A & 4.575 & 3.975--5.225 & 4.407--4.993  & 2.75 \\ 
    B & 5.625 & 5.025--6.275 & 5.457--6.043  & 2.25 \\ 
    C & 6.125 & 5.525--6.775 & 5.957--6.543  & 2.05 \\ 
    D & 7.175 & 6.575--7.825 & 7.007--7.593  & 1.75 \\
\hline
\end{tabular}

%% file: tables/tab2a_FlagSummaryResults_kuband.tex
\begin{tabular}{c|c|c|c|c}
\hline
SPW &     Channels & Scans \textgreater1\% & From SK & From NSR \\
    & Flagged (\%) &          Flagged (\%) &  Cut (\%) &  Cut (\%) \\
\hline
  B &         0.07 &                  0.79 &  81.75 &   45.69 \\
  C &         0.06 &                  0.79 &  75.23 &   36.77 \\
  D &         0.10 &                  0.79 &  66.42 &   61.37 \\
\hline
\end{tabular}

%% file: tables/tab2b_FlagSummaryResults_cband.tex
\begin{tabular}{c|c|c|c|c}
\hline
SPW &     Channels & Scans \textgreater1\% & From SK & From NSR \\
    & Flagged (\%) &          Flagged (\%) &  Cut (\%) &  Cut (\%) \\
\hline
  A &         2.12 &                  9.09 &  89.32 &   64.07 \\
  B &         0.22 &                  2.02 &  92.10 &   16.58 \\
  C &         0.89 &                 29.29 &  98.23 &   55.62 \\
  D &         1.42 &                  8.08 &  89.82 &   50.05 \\
\hline
\end{tabular}

%% file: tables/tab3_dataSetSummary.tex
\begin{tabular}{l|c|c|c|c|c}
\hline
    Experiment & Frequency & Beam FWHM & Aperture Flux Density & References \\
               & (GHz) & (arcmin) & (Jy) &  \\
\hline
    CGPS (DRAO/Haslam)          & 0.408 & 2.8  &  $16\pm3$     & 
    \cite{taylor_canadian_2003, tung_high_2017}\\
    CGPS (DRAO/Stockert)*       & 1.42  & 1    &  ...  & \cite{taylor_canadian_2003, landecker_survey_2010}\\
    Reich/Stockert              & 1.42  & 36.0  & $17\pm2$ & \cite{reich_radio_1982}\\
    GBT C-band (SPW A)         & 4.575 & 2.75  &  $19\pm2$ & \cite{abitbol_constraining_2018}\\
    GBT C-band (SPW B)         & 5.625 & 2.24  & $20\pm2$ & `` ''             \\
    GBT C-band (SPW C)         & 6.125 & 2.05  & $19\pm2$ & `` ''             \\
    GBT Ku-band (SPW B)        & 13.1  & 0.96  & $20\pm3$ & This work         \\
    QUIJOTE$^+$                & 11 & 55.38 & $21.8\pm1.2$ & \cite{rubino-martin_quijote_2023} \\
    QUIJOTE$^+$                & 13 & 55.84 & $23.4\pm1.2$& ``'' \\
    QUIJOTE                    & 17 & 38.95 & $25.8\pm1.4$ & ``'' \\
    QUIJOTE                    & 19 & 40.32 & $26.6\pm1.5$ & ``'' \\
    \textit{Planck LFI}             & 28.4 & 32.3  & $27.8\pm0.7$ & \cite{planck_results_2016a}\\
    \textit{Planck LFI}             & 44.1 & 27.1  & $25.3\pm0.8$ & `` ''             \\
    \textit{Planck LFI}             & 70.4 & 13.3  & $24.7\pm1.1$ & `` ''             \\
    \textit{Planck HFI}$^\dagger$   & 100  & 9.7   & $51\pm2$   & `` ''             \\
    \textit{Planck HFI}             & 143  & 7.3   & $84\pm4$ & `` ''             \\
    \textit{Planck HFI}$^\dagger$   & 217  & 5.0   & $362\pm12$   & `` ''             \\
    \textit{Planck HFI}             & 353  & 4.8   & $1,460\pm50$ & `` ''             \\
    \textit{Planck HFI}             & 545  & 4.7   & $5,100\pm200$ & `` ''             \\
    \textit{Planck} HFI             & 857  & 4.3   & $17,400\pm500$ & `` ''             \\
        \textit{COBE}-DIRBE     & 1,249 & 39.5  & $38,000\pm1,000$ & \cite{hauser_cobe_1998}    \\
    \textit{COBE}-DIRBE         & 2,141 & 40.4  & $66,000\pm2,000$ & `` ''             \\
    \textit{COBE}-DIRBE         & 2,997 & 41.0  & $43,000\pm1,000$ & `` ''             \\
    IRIS (100 $\mu$m)* & 3,000  & 4.3   & ...   &  \cite{miville-deschenes_iris_2005}   \\
    IRIS (60 $\mu$m)*  & 5,000  & 4.0   & ...   & `` ''             \\
    IRIS (25 $\mu$m)*  & 12,000 & 3.8   & ...   & `` ''             \\
    IRIS (12 $\mu$m)*  & 25,000 & 3.8   & ...   & `` ''             \\
    SPHEREx* (3.4 $\mu$m)   &  88,000 & 372 & ... &  
   \cite{spherex_team_spherex_2025}, \cite{cukierman_spectral_2026} \\
    SPHEREx* (3.3 $\mu$m)   & 90,000 & 372 & ... &  ``''\\
\hline
\end{tabular}

%% file: tables/tab4_modelParamSummary.tex
\begin{tabular}{c|c|c|c}
        \hline
            Foreground & Spectral Radiance Model & Free Parameters & Additional Information \\
                       & [Jy sr$^{-1}$] &  &  \\
            \hline
            \hline
            Conversion Factor & $\frac{2 k_\mathrm{b} \nu^2}{c^2}  \times 10^{26}$ & ... & from K to Jy sr$^{-1}$ \\
            \hline
            Thermal Dust    & $x(\nu)=\frac{h \nu}{k T_\mathrm{D}}$ & $A_\mathrm{D}$ [K]   & $\nu_0 = 545$ GHz \\
                        & $I_\mathrm{D}(\nu) = A_\mathrm{D} \left(\frac{x}{x_0}\right)^{\beta_\mathrm{D} + 1} \frac{e^{x_0}-1}{e^x-1} \times I_\mathrm{RJ}$ & $\beta_\mathrm{D}$, $T_\mathrm{D}$ [K]   &   \\
            \hline
                    & $g_\mathrm{FF}(\nu)=\text{ln}\left(\frac{0.04955}{\nu/10^9}\right)+1.5\text{ln}(T_\mathrm{e})$ &  &  \\
            Free-Free & $T_\mathrm{FF}(\nu) = 0.03014 \frac{T^{-0.15}}{(\nu/10^9)^2}\times \text{EM} \times g_\mathrm{FF}(\nu)$ & EM [cm$^{-6}$ pc] & $T_e = 8000$ K \\
                    & $I_\mathrm{FF}(\nu) = T_e (1-e^{-T_\mathrm{FF}(\nu)}) \times I_\mathrm{RJ}$ & & \\            
            \hline
            Optically thick free-free & $I_\text{UCHII}(\nu) = I_\text{FF}(\nu) \times \pi(\theta_\text{UCHII}/206,265)^2$ & $\theta_\text{UCHII}$ [arcsec], EM$_\text{UCHII}$ [cm$^{-6}$ pc] & \\
            \hline
                Spinning Dust & $S_\mathrm{SD}(\nu) = A_\mathrm{SD} \times \mathrm{exp} \left(-\frac{1}{2} \left(\frac{\ln{(\nu/\nu_\mathrm{SD})}}{W_\mathrm{SD}}\right)^2\right)$  & $A_\mathrm{SD}$ [Jy], $\nu_\mathrm{SD}$ [GHz] &  $W_\mathrm{SD} = 0.7$\\
            \hline
                & $X = \frac{h \nu}{k_\mathrm{b} T_\mathrm{CMB}}$ &   &   \\
            CMB & $g_\mathrm{f}(\nu) = \frac{(e^X-1)^2}{X^2 e^X}$& $A_\mathrm{CMB}$ [K] & $T_\mathrm{CMB} = 2.7255$ K  \\
                & $I_\mathrm{CMB}(\nu) = A_\mathrm{CMB}/g_\mathrm{f}(\nu) \times I_\mathrm{RJ}$ &   &   \\
        \hline    
\end{tabular}

%% file: tables/tab5_mcmcSamplingResults.tex
\begin{tabular}{c|ccccccc|cc}
    \hline
        & EM$_\text{diffuse}$ & $A_\text{SD}$ & $\nu_\text{SD}$ & $A_\text{D}$ & $\beta_\text{D}$ & $T_\text{D}$ & $A_\text{CMB}$ & WAIC & $p_\text{WAIC}$\\
         \textbf{Spinning}  & (cm$^{-6}$ pc) & (Jy) &  (GHz)  & ($\mu$K) & (--) & (K) & ($\mu$K) & & \\
        \textbf{Dust} & $332^{+20}_{-20}$ & $14.1^{+1.1}_{-1.1}$ & $27^{+2}_{-2}$ & $1068^{18}_{-18}$ & $1.64^{+0.05}_{-0.05}$ & $21.6^{+0.4}_{-0.4}$ & $33^{+22}_{-23}$ & 38 & 8 \\  
    \hline
        & EM$_\text{diffuse}$ & EM$_\text{UCHII}$ & $\theta_\text{UCHII}$ & $A_\text{D}$ & $\beta_\text{D}$ & $T_\text{D}$ & $A_\text{CMB}$ & WAIC & $p_\text{WAIC}$ \\
        & (cm$^{-6}$ pc) & (10$^9$cm$^{-6}$ pc) & (arcsec) & ($\mu$K) & (--) & (K) & ($\mu$K) &  & \\
        \textbf{UC-HII} & $307^{+26}_{-28}$ & $0.5^{+0.3}_{-0.2}$ & $2.4^{+0.6}_{-0.4}$ & $1081^{+17}_{-17}$ & $1.59^{+0.05}_{-0.05}$ & $22.0^{+0.4}_{-0.4}$ & $-74^{+24}_{-25}$ & 47 & 12 \\
    \hline
\end{tabular}